\documentclass[prd,preprint,longbibliography,
  showpacs,showkeys,lengthcheck,
  nofootinbib,tightenlines,onecolumn,notitlepage,
  preprintnumbers,superscriptaddress]{revtex4-1}

\usepackage{graphicx}
\usepackage[usenames,dvipsnames]{xcolor}
\graphicspath{{figures/}}

\usepackage{hyperref}
\hypersetup{colorlinks=true,citecolor=blue,linkcolor=blue,urlcolor=blue}

\usepackage{diagbox}
\usepackage{booktabs}

\usepackage{amsmath,amssymb,amsfonts}
\usepackage{mathrsfs}
\usepackage{slashed}
\usepackage{bm}
\usepackage{bbm}

\usepackage[final]{changes} 

\usepackage[utf8]{inputenc}
\usepackage{newtxtext}
\usepackage[mathcal]{euscript}

\renewcommand*{\d}{\mathop{}\!\mathrm{d}}
\newcommand*{\e}{\mathop{}\!\mathrm{e}}
\renewcommand*{\i}{\mathop{}\!\mathrm{i}}

\begin{document}

\title{Light front synchronization and rest frame densities of the proton: Electromagnetic densities}

\author{Adam Freese}
\email{afreese@uw.edu}
\address{Department of Physics, University of Washington, Seattle, WA 98195, USA}

\author{Gerald A. Miller}
\email{miller@uw.edu}
\address{Department of Physics, University of Washington, Seattle, WA 98195, USA}

\begin{abstract}
  We clarify the physical origin and meaning of the two-dimensional relativistic
  densities of the light front formalism.
  The densities are shown to originate entirely from the use of light front time
  instead of instant form time,
  which physically corresponds to using an alternative synchronization convention.
  This is shown by using tilted light front coordinates,
  which consist of light front time and ordinary spatial coordinates,
  and which are also used to show that
  the obtained densities describe a system at rest
  rather than at infinite momentum.
  These coordinates allow all four components of the
  electromagnetic current density to be given clear physical meanings.
  We explicate the formalism for spin-half targets,
  obtaining charge and current densities of the proton and neutron
  using empirical form factor parametrizations,
  as well as up and down quark densities and currents.
  Angular modulations in the densities of transversely-polarized states are
  explained as originating from redshifts and blueshifts due to quarks moving
  in different longitudinal directions.
\end{abstract}

\preprint{NT@UW-23-03}

\maketitle


\section{Introduction}
\label{sec:intro}

There has recently been renewed interest and debate about the proper manner
of understanding spatial densities of hadrons,
especially regarding their application to the energy-momentum
tensor~\cite{Polyakov:2018zvc,Lorce:2018egm,Freese:2021czn,Panteleeva:2022uii}.
This interest is especially pertinent to the upcoming
Electron Ion Collider~\cite{Boer:2011fh,Accardi:2012qut,AbdulKhalek:2021gbh},
since spatial tomography of partons in hadrons
is one of the facility's central focuses.
It is vital to the success of this program that an accurate formalism
be employed to obtain spatial distributions of partons within hadrons.

For the longest time, most authors have calculated relativistic densities
using three-dimensional Fourier transforms of form factors,
obtaining results referred to as Breit frame densities or
Sachs densities~\cite{Sachs:1962zzc}.
For a long time, too, this approach has been criticized as relativistically
inexact and as not following from elementary field-theoretic
definitions~\cite{Fleming:1974af,Burkardt:2000za,Miller:2018ybm}.
While most authors continue to use Breit frame densities uncritically,
a variety of criticisms and defenses of the Breit frame densities
and alternative formalisms have been proposed
(see Refs.~\cite{Fleming:1974af,Burkardt:2000za,Miller:2018ybm,Lorce:2018egm,Epelbaum:2022fjc,Li:2022ldb,Chen:2022smg,Freese:2022fat} for a variety of perspectives).

The light front formalism has stood out in these debates as providing
relativistically exact two-dimensional densities~\cite{Burkardt:2002hr,Miller:2010nz}
that can be obtained from elementary field-theoretic definitions~\cite{Burkardt:2000za,Freese:2021czn}
in a wave-packet-independent way~\cite{Freese:2022fat}.
Additionally, the light front densities are related to
generalized parton distributions~\cite{Diehl:2002he,Burkardt:2002hr,Miller:2018ybm},
which simultaneously encode electromagnetic and gravitational form factors
through their polynomiality property~\cite{Ji:1998pc,Diehl:2003ny},
and which extend light front densities to three-dimensional distributions
over the transverse plane and the light front momentum fraction $x$.

Despite the success of the light front formalism,
misgivings about light front densities persist.
It is widely believed that light front densities describe the apparent
structure of a system moving at the speed of light,
and thus contain kinematic distortions caused by boosting
to infinite momentum.
Another criticism---given in Ref.~\cite{Chen:2022smg}
for instance---is that the ``minus'' components of four-vector
and tensor densities do not have clear meanings,
and indeed existing works on light front densities merely ignore these components.

Both of these misgivings can be addressed with a
minor change to the light front coordinates.
By using light front time $x^+=t+z$
with ordinary spatial coordinates $(x,y,z)$---an
idea first explored in Ref.~\cite{Blunden:1999wb},
where these are called \emph{tilted light front coordinates}---all
existing results for light front densities are reproduced,
but the previously-neglected components of vectors and tensors
obtain clear physical meanings.
Moreover, by using ordinary spatial coordinates,
it can unambiguously be shown that the
previously-known and newly-found light front densities alike
describe a system moving at any velocity,
even a system at rest.

The use of tilted light front coordinates has
a straightforward conceptual interpretation:
using light front time $x^+$ instead of instant form time $t$ simply
means changing the rules for how spatially distant clocks are synchronized.
In this work, we will show how the entire formalism of light front densities
follows from changing the time synchronization convention.
Due to the lengthy exposition required for the coordinate system,
this paper will focus entirely on electromagnetic densities of spin-half systems;
the energy-momentum tensor will be explored in a sequel work.

This work is organized as follows.
Tilted coordinates are defined in Sec.~\ref{sec:tilted},
which also contains a detailed exposition of their properties:
general properties are explored in Sec.~\ref{sec:properties};
Lorentz transforms in Sec.~\ref{sec:lorentz};
conserved currents and identification of the
``charge'' component in Sec.~\ref{sec:cc}.
The wave packet and spin density matrix are constructed in Sec.~\ref{sec:packet}.
The formulas for intrinsic electromagnetic densities
of spin-half systems are derived in Sec.~\ref{sec:theory}.
In Sec.~\ref{sec:empirical}, these formulas are applied
to empirical proton and neutron form factors
to obtain charge and current densities,
as well as up and down quark densities.
We provide concluding remarks and an outlook in Sec.~\ref{sec:end}.
A uniqueness proof for light front synchronization
is given in Appendix~\ref{sec:unique}.


\section{Tilted coordinates and light front synchronization}
\label{sec:tilted}

Tilted light front coordinates (hereafter ``tilted coordinates'')
were first defined by Blunden, Burkardt and Miller (BBM)
in Ref.~\cite{Blunden:1999wb}\footnote{
  BBM use $\zeta = -z$ instead of $\tilde{z}=z$ as a coordinate,
  but this is a minor difference.
}:
\begin{subequations}
  \label{eqn:tilted}
  \begin{align}
    \tilde{\tau} = \tilde{x}^0 &\equiv x^+ = t + z \\
    \tilde{x}^1 &\equiv x \\
    \tilde{x}^2 &\equiv y \\
    \tilde{x}^3 &\equiv z
    \,.
  \end{align}
\end{subequations}
We use tildes for quantities in tilted coordinates for clarity.
The three spatial coordinates are the usual Cartesian coordinates
of Euclidean space.
Crucially, this means that a system at rest in tilted coordinates
is really at rest as understood by common sense.

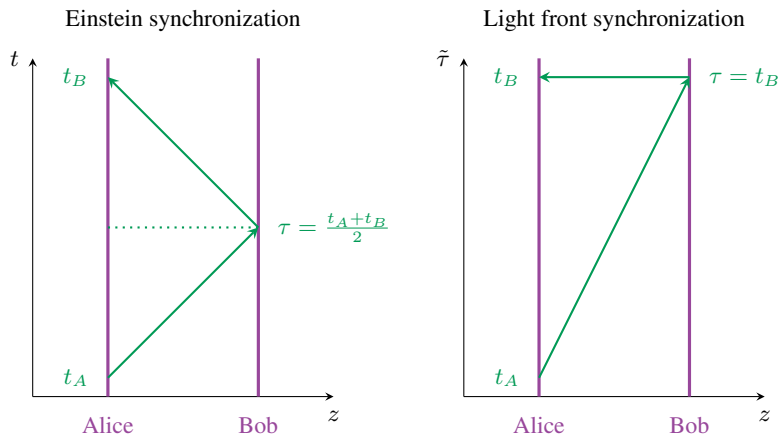
\begin{figure}
  \centering
  \begin{tikzpicture}
    \draw[-stealth] (0,-0.25) -- (0, 4.25) node[left=2pt]  {$t$} ;
    \draw[-stealth] (0,-0.25) -- (4,-0.25) node[below=2pt] {$z$};
    \draw[very thick, Purple] (1,-0.25) node[below=4pt] {Alice} -- (1,4.25);
    \draw[very thick, Purple] (3,-0.25) node[below=4pt] {Bob}   -- (3,4.25);
    \draw[-stealth, thick, ForestGreen] (1,0) node[left=4pt] {$t_A$} -- (3,2);
    \draw[-stealth, thick, ForestGreen] (3,2) -- (1,4) node[left=4pt] {$t_B$};
    \draw[dotted, thick, ForestGreen] (1,2) -- (3,2) node[right=4pt] {$\tau=\frac{t_A+t_B}{2}$};
    \node at (2,4.75) {Einstein synchronization};
  \end{tikzpicture}
  ~~~
  \begin{tikzpicture}
    \draw[-stealth] (0,-0.25) -- (0, 4.25) node[left=2pt]  {$\tilde{\tau}$} ;
    \draw[-stealth] (0,-0.25) -- (4,-0.25) node[below=2pt] {$z$};
    \draw[very thick, Purple] (1,-0.25) node[below=4pt] {Alice} -- (1,4.25);
    \draw[very thick, Purple] (3,-0.25) node[below=4pt] {Bob}   -- (3,4.25);
    \draw[-stealth, thick, ForestGreen] (1,0) node[left=4pt] {$t_A$} -- (3,4) node[right=4pt] {$\tau=t_B$};
    \draw[-stealth, thick, ForestGreen] (3,4) -- (1,4) node[left=4pt] {$t_B$};
    \node at (2,4.75) {Light front synchronization};
  \end{tikzpicture}
  \caption{
    Pictorial representations of time synchronization conventions.
    In these pictures, Alice sends a light signal to Bob,
    who promptly sends a return signal.
    The time $\Delta t = t_B - t_A$ elapsed throughout the exchange
    from Alice's perspective is a local observable, independent of synchronization convention.
    In the Einstein convention (left panel),
    the local time measured by Bob's clock when the signal is reflected
    is halfway between the start and end times.
    In light front synchronization (right panel),
    the clocks are synchronized under the convention that the return signal
    travels instantaneously,
    so Bob's clock reads $t_B$ when the signal is reflected.
    The invariance of $\Delta t$ requires that the signal to Bob travel at $c/2$
    in this convention.
  }
  \label{fig:synchronize}
\end{figure}

The tilted time variable $\tilde{\tau}$  is the same as light front time,
and can be given an operational meaning.
In fact, the familiar time $t$ of Minkowski coordinates is also defined
operationally, as first clarified by Einstein~\cite{Einstein:1905ve}.
The local time $\tau$ at which an event occurs is determined by its simultaneity
with the reading of a clock in the local vicinity of the event.
Einstein postulates \emph{by definition}
(``\emph{durch Definition}''---his emphasis)
that distant clocks be synchronized by light signals under
the convention that the time delay of the signal is equal in both directions.
This definition has come to be called the Einstein synchronization convention,
and using it gives us the Einstein (or instant form) time coordinate $\tau=x^0=t$.
A pictorial representation of this convention is given in the left panel of
Fig.~\ref{fig:synchronize}.

The use of Einstein's convention is ultimately arbitrary---a point
which has been discussed for a long time by both physicists
(see Refs.~\cite{Zhang:1995test,Anderson:1998mu} and references therein)
and philosophers~\cite{reichenbach2012philosophy,gruenbaum2012philosophical}\footnote{
  This feature of special relativity has also recently been popularized
  by the science education channel Veritasium~\cite{Veritasium:2020oct}.
}.
The use of an anisotropic synchronization convention notoriously results
in the same empirical predictions as Einstein's convention,
since mathematically this merely corresponds to a
coordinate transformation~\cite{Anderson:1998mu}.

Light front time (and thus tilted time) is defined
using an anisotropic synchronization convention,
which we call light front synchronization.
Observers aligned in the $z$ direction synchronize their clocks
\emph{under the convention that} light travels instantaneously in the $-z$ direction.
For instance, if Alice is located downstream of Bob
($z_{\text{Alice}} < z_{\text{Bob}}$),
she synchronizes her clock to read the time she currently sees on Bob's clock.
This is depicted in the right panel of Fig.~\ref{fig:synchronize}.
The local time elapsed during the signal's round trip is convention-independent,
meaning the two-way speed of light is given by the invariant constant
$c$~\cite{Zhang:1995test,reichenbach2012philosophy}.
Under light front synchronization,
this means light must travel at $c/2$ in the $+z$ direction,
as depicted in the right panel of Fig.~\ref{fig:synchronize}.
Clocks at the same $z$ coordinate but displaced in the transverse $xy$ plane
are synchronized using the Einstein convention.
Following these rules for clock synchronization
gives us the light front time $\tau=\tilde{x}^0=x^+$.

As noted above, the empirical predictions of a relativistic theory are
independent of the synchronization convention.
The use of a particular convention is ultimately arbitrary,
but different conventions may be more conducive to solving particular problems.
The motivation for using light front synchronization---and thus tilted
coordinates---is that light front time is invariant under
a Galilean subgroup of the Poincar\'{e} group~\cite{Soper:1972xc,Burkardt:2002hr}.
Ultimately, this allows physical densities at fixed $\tilde{\tau}$ to be
separated into state-independent internal structures and state-dependent
smearing functions in the manner explored in Ref.~\cite{Freese:2022fat}.


\subsection{Properties of tilted coordinates}
\label{sec:properties}

Let us consider some of the properties of tilted coordinates,
especially since they differ from the more familiar
light front coordinates.
Some of these properties were reported previously by
BBM~\cite{Blunden:1999wb},
but some of these relationships are new.
Firstly, the metric and inverse metric are given by:
\begin{subequations}
  \begin{align}
    \tilde{g}_{\mu\nu}
    &=
    \frac{\partial x^\alpha}{\partial \tilde{x}^\mu}
    \frac{\partial x^\beta }{\partial \tilde{x}^\nu}
    g_{\alpha\beta}
    =
    \left[
      \begin{array}{cccc}
        \phantom{-}1 & \phantom{-}0 & \phantom{-}0 & -1 \\
        \phantom{-}0 & -1 & \phantom{-}0 & \phantom{-}0 \\
        \phantom{-}0 & \phantom{-}0 & -1 & \phantom{-}0 \\
        -1 & \phantom{-}0 & \phantom{-}0 & \phantom{-}0
      \end{array}
      \right]
    \\
    \tilde{g}^{\mu\nu}
    &=
    \frac{\partial \tilde{x}^\mu}{\partial x^\alpha}
    \frac{\partial \tilde{x}^\nu}{\partial x^\beta }
    g^{\alpha\beta}
    =
    \left[
      \begin{array}{cccc}
        \phantom{-}0 & \phantom{-}0 & \phantom{-}0 & -1 \\
        \phantom{-}0 & -1 & \phantom{-}0 & \phantom{-}0 \\
        \phantom{-}0 & \phantom{-}0 & -1 & \phantom{-}0 \\
        -1 & \phantom{-}0 & \phantom{-}0 & -1
      \end{array}
      \right]
  \end{align}
\end{subequations}
where $g_{\mu\nu} = g^{\mu\nu} = \mathrm{diag}(+1,-1,-1,-1)$
is the usual Minkowski metric\footnote{
  Note that, while using tilted coordinates, we use numerical indices
  $0,1,2,3$ rather than $+,1,2,3$ to label components of tensors.
}.
This means that the invariant proper time element is given by:
\begin{align}
  \d s^2
  =
  \tilde{g}_{\mu\nu} \d\tilde{x}^\mu \d\tilde{x}^\nu
  =
  \d\tilde{\tau}^2
  -
  2\d\tilde{\tau}\d\tilde{z}
  -
  \d\tilde{\bm{x}}_\perp^2
  \,,
\end{align}
where $\tilde{\bm{x}}_\perp = (\tilde{x},\tilde{y})$ are the transverse coordinates,
while the d'Alembertian is given by:
\begin{align}
  \partial^2
  =
  \tilde{g}^{\mu\nu}
  \tilde{\partial}_\mu
  \tilde{\partial}_\nu
  =
  -2 \tilde{\partial}_z \tilde{\partial}_\tau
  -
  \tilde{\bm{\nabla}}^2
  \,,
\end{align}
where $\tilde{\bm{\nabla}} = (\tilde{\partial}_x,\tilde{\partial}_y,\tilde{\partial}_z)$
is the three-dimensional gradient.
The latter was found previously by BBM~\cite{Blunden:1999wb}.

Covariant components of a four-vector $\tilde{A}_\mu$ (lower index)
are related to contravariant components $\tilde{A}^\mu$ (upper index)
through the metric:
\begin{subequations}
  \label{eqn:covcon}
  \begin{align}
    &
    \tilde{A}_\mu = \tilde{g}_{\mu\nu} \tilde{A}^\nu
    &
    \tilde{A}^\mu = \tilde{g}^{\mu\nu} \tilde{A}_\nu
    \,,
  \end{align}
  which in terms of individual components gives:
  \begin{align}
    &
    \tilde{A}_0
    =
    \tilde{A}^0 - \tilde{A}^3
    &
    \tilde{A}^0
    =
    -
    \tilde{A}_3
    \,\phantom{.}
    \\
    &
    \tilde{A}_1
    =
    -\tilde{A}^1
    &
    \tilde{A}^1
    =
    -\tilde{A}_1
    \,\phantom{.}
    \\
    &
    \tilde{A}_2
    =
    -\tilde{A}^2
    &
    \tilde{A}^2
    =
    -\tilde{A}_2
    \,\phantom{.}
    \\
    &
    \tilde{A}_3
    =
    -\tilde{A}^0
    &
    \tilde{A}^3
    =
    -\tilde{A}_0 - \tilde{A}_3
    \,.
  \end{align}
\end{subequations}

Components of four-momentum are defined as generators of spacetime translations:
\begin{align}
  i[\hat{P}_\mu, \hat{O}(x)]
  =
  \partial_\mu \hat{O}(x)
  \,,
\end{align}
and accordingly the energy and vector momentum are identified using
the covariant vector $\tilde{p}_\mu$:
\begin{align}
  \tilde{p}_\mu
  \equiv
  (\tilde{E};-\tilde{p}_x,-\tilde{p}_y,-\tilde{p}_z)
  \,.
\end{align}
Note that when we use $x$, $y$ or $z$ as a subscript,
we are indicating a component of the three-vector $\tilde{\bm{p}}$
rather than components of the four-vector $p_\mu$:
\begin{align}
  \tilde{\bm{p}}
  =
  (\tilde{p}_x, \tilde{p}_y, \tilde{p}_z)
  =
  (-\tilde{p}_1, -\tilde{p}_2, -\tilde{p}_3)
  \,.
\end{align}
With this identification, the tilted energy and vector momentum
become time and space translation generators as usual:
\begin{subequations}
  \begin{align}
    i[\tilde{E},\hat{O}(x)]
    &=
    \tilde{\partial}_0 \hat{O}(x)
    \\
    -i[\tilde{\bm{p}},\hat{O}(x)]
    &=
    \tilde{\bm{\nabla}} \hat{O}(x)
    \,.
  \end{align}
\end{subequations}
The scalar product between four-momentum and a spacetime coordinate is:
\begin{align}
  \tilde{p}\cdot\tilde{x}
  =
  \tilde{p}_\mu \tilde{x}^\mu
  =
  \tilde{E} \tilde{\tau}
  -
  \tilde{\bm{p}}\cdot\tilde{\bm{x}}
  \,,
\end{align}
which is a point of central importance to the density formalism.
The mass-shell relation can be written:
\begin{subequations}
  \begin{align}
    m^2
    &=
    \tilde{g}^{\mu\nu} \tilde{p}_\mu \tilde{p}_\nu
    =
    2 \tilde{p}_z \tilde{E}
    -
    \tilde{\bm{p}}^2
    \\
    \label{eqn:energy}
    \tilde{E}
    &=
    \frac{m^2 + \tilde{\bm{p}}^2}{2 \tilde{p}_z}
    \,,
  \end{align}
\end{subequations}
which was found previously by BBM~\cite{Blunden:1999wb}.

The relationships between the tilted and Minkowski four-momenta
can be found using:
\begin{align}
  \tilde{p}_\mu
  =
  \frac{\partial x^\nu}{\partial\tilde{x}^\mu}
  p_\nu
  \,,
\end{align}
which entails:
\begin{subequations}
  \begin{align}
    \tilde{E}
    &=
    E
    \\
    \tilde{p}_x
    &=
    p_x
    \\
    \tilde{p}_y
    &=
    p_y
    \\
    \tilde{p}_z
    &=
    E + p_z
    =
    p^+
    \,.
  \end{align}
\end{subequations}
Two remarkable things have happened here.
Firstly, the tilted energy is equal to the instant form energy.
This means that a mass decomposition in tilted coordinates
will be the same as a mass decomposition in Minkowski coordinates
(in contrast to the standard light front energy $p^-$~\cite{Lorce:2021xku}),
and a spatial density of $\tilde{E}$ is a density of
the familiar instant form energy $E$.
Secondly, the conjugate momentum to the $z$ coordinate is $p^+$,
which is the momentum variable used to define momentum fractions in
parton distribution functions and generalized parton distributions.

Lastly, we discuss particle velocities
and the conditions for a particle being at rest.
Classically, the velocity of a free particle
is obtained from Hamilton's equations:
\begin{subequations}
  \label{eqn:velocity}
  \begin{align}
    \tilde{v}_{a}
    =
    \frac{\partial\tilde{E}}{\partial\tilde{p}_{a}}
    \,,
  \end{align}
  (where $a=x,y,z$)
  which, upon using Eq.~(\ref{eqn:energy}) gives
  \begin{align}
    \tilde{v}_x
    &=
    \frac{\tilde{p}_x}{\tilde{p}_z}
    \\
    \tilde{v}_y
    &=
    \frac{\tilde{p}_y}{\tilde{p}_z}
    \\
    \tilde{v}_z
    &=
    1
    -
    \frac{\tilde{E}}{\tilde{p}_z}
  \end{align}
  for the individual components of the velocity.
  All three of these can be summarized with the help of Eq.~(\ref{eqn:covcon}) as:
  \begin{align}
    \tilde{v}^i
    =
    \frac{\tilde{p}^i}{\tilde{p}_z}
    \,.
  \end{align}
\end{subequations}
Note the upper index in the components of the momentum four-vector here.
This relationship can be made quantum mechanical simply
by promoting it to an operator relationship.
The free particle is at rest when $\tilde{\bm{v}}=0$,
which occurs when:
\begin{subequations}
  \label{eqn:rest}
  \begin{align}
    \tilde{p}_x &= 0 \\
    \tilde{p}_y &= 0 \\
    \tilde{p}_z &= m
    \,.
  \end{align}
\end{subequations}
The relationship between longitudinal velocity and longitudinal momentum
is strange and counter-intuitive,
but follows from the elementary definition of momentum
as the generator of spatial translations \emph{at a fixed time}
(which is fixed light front time here).
Regardless, Eq.~(\ref{eqn:rest}) allows us to clearly identify
if a particle is at rest---and since the spatial coordinates
$\tilde{\bm{x}} = \bm{x}$ are the usual Cartesian coordinates,
a system at rest with respect to these coordinates is unambiguously
at rest as understood by common sense.


\subsection{Lorentz transforms and the Galilean subgroup}
\label{sec:lorentz}

We next explore the algebra of the Lorentz group
and the effects of finite boosts in tilted coordinates.
Since the goal of this work is to obtain rest frame densities,
special focus will be given to properties that are invariant under boosts,
since these properties remain unaltered when boosting the system to rest.

The standard generator basis for the Poincar\'{e} group in Minkowski coordinates
consists of four translation generators $P_\mu$
and six Lorentz transform generators $J^{\mu\nu} = -J^{\nu\mu}$,
with the latter consisting of
three rotation generators $\bm{J} = (J^{23}, J^{31}, J^{12})$
and three boost generators $\bm{K} = (J^{10}, J^{20}, J^{30})$.
We have already discussed the translation generators in tilted coordinates above.
The tilted rotation generators are the same as in instant form coordinates,
$\tilde{\bm{J}} = \bm{J}$, since the same spatial coordinates
are used in both systems.
The boosts in tilted coordinates remain to be analyzed.

Under an \emph{infinitesimal} passive transverse \emph{instant form} boost,
with velocities $\d\beta_x$ and $\d\beta_y$ in the $x$ and $y$ directions,
the instant form coordinates transform as:
\begin{subequations}
  \begin{align}
    t'
    &=
    t - \d\beta_x x - \d\beta_y y
    \\
    x'
    &=
    x - \d\beta_x t
    \\
    y'
    &=
    y - \d\beta_y t
    \\
    z'
    &=
    z
    \,.
  \end{align}
\end{subequations}
In terms of tilted coordinates, this same infinitesimal transformation can be written:
\begin{subequations}
  \begin{align}
    \tilde{\tau}'
    &=
    \tilde{\tau} - \d\beta_x \tilde{x} - \d\beta_y \tilde{y}
    \\
    \tilde{x}'
    &=
    \tilde{x} - \d\beta_x \tilde{\tau} + \d\beta_x \tilde{z}
    \\
    \tilde{y}'
    &=
    \tilde{y} - \d\beta_y \tilde{\tau} + \d\beta_y \tilde{z}
    \\
    \tilde{z}'
    &=
    \tilde{z}
    \,.
  \end{align}
\end{subequations}
Remarkably, this transformation is not merely a boost in tilted coordinates,
but a combined boost and rotation.
In fact, this is the origin of Terrell-Penrose rotations~\cite{Penrose:1959vz,Terrell:1959zz}:
what is merely boost according to a spatially extended reference frame
with clocks synchronized by the Einstein convention
\emph{looks like} a boost and a rotation,
and in fact \emph{is} a boost and a rotation under the light front synchronization convention.
Because of this, to define a mere boost in tilted coordinates,
we must counteract the Terrell-Penrose rotation.
The appropriate boost generators are:
\begin{subequations}
  \begin{align}
    \tilde{B}_1
    &=
    K_1 - J_2
    \\
    \tilde{B}_2
    &=
    K_2 + J_1
    \,.
  \end{align}
  The remaining boost generator remains unmodified:
  \begin{align}
    \tilde{B}_3
    &=
    K_3
    \,.
  \end{align}
\end{subequations}
All three of these boosts modify the $\tilde{z}$ coordinate in unintuitive ways.
Under a finite \emph{light front} transverse boost by a velocity $v$
along the $x$ direction, the tilted coordinates transform as:
\begin{subequations}
  \label{eqn:boost:transverse}
  \begin{align}
    \tilde{\tau}'
    &=
    \tilde{\tau}
    \\
    \tilde{x}'
    &=
    \tilde{x}
    -
    v\tilde{\tau}
    \\
    \tilde{y}'
    &=
    \tilde{y}
    \\
    \tilde{z}'
    &=
    \tilde{z}
    +
    v \tilde{x}
    -
    \frac{v^2}{2} \tilde{\tau}
    \,,
  \end{align}
\end{subequations}
while under a finite longitudinal boost of rapidity\footnote{
  Recall that in instant form coordinates,
  the rapidity of a boost is given by
  $\eta = \frac{1}{2}\log\left(\frac{c+v_{\text{IF}}}{c-v_{\text{IF}}}\right)$.
  In tilted coordinates, Eq.~(\ref{eqn:rapidity}) instead gives the
  relationship between rapidity and longitudinal velocity.
} $\eta$ they transform as:
\begin{subequations}
  \label{eqn:boost:longitudinal}
  \begin{align}
    \tilde{\tau}'
    &=
    \e^{-\eta} \tilde{\tau}
    \\
    \tilde{x}'
    &=
    \tilde{x}
    \\
    \tilde{y}'
    &=
    \tilde{y}
    \\
    \tilde{z}'
    &=
    \e^{\eta}
    \tilde{z}
    -
    \sinh(\eta) \tilde{\tau}
    \,.
  \end{align}
\end{subequations}
An object that is stationary in the original frame moves
with a velocity
\begin{align}
  \label{eqn:rapidity}
  v_z
  =
  -
  \e^\eta \sinh(\eta)
  =
  \frac{1}{2}\big( 1 - \e^{2\eta} \big)
\end{align}
in the new frame,
which is bounded to $(-\infty, 1/2)$---the bounds being
the one-way speed of light in each direction along the $z$ axis.

The longitudinal boost is worth understanding conceptually.
Let $\Delta\tilde{\tau}$ be a time interval measured by a clock in the original frame;
the same time interval according to the transformed frame will be
$\Delta\tilde{\tau}' = \e^{-\eta} \Delta{\tilde{\tau}}$.
Since the boost was a passive transformation,
the now-moving clock will measure the interval
$\Delta\tilde{\tau} = \e^\eta\Delta\tilde{\tau}'$
while a stationary clock in the new frame measures $\Delta\tilde{\tau}'$.
Thus, when $\eta > 0$, time measured by the moving clock is dilated,
while it is instead quickened when $\eta < 0$.

Next, let $\Delta\tilde{z}$ be the length between two comoving bodies as measured
in the original frame;
the length will be $\Delta\tilde{z}' = \e^{\eta}\Delta\tilde{z}$ in the transformed frame.
If $\eta > 0$, the length will appear dilated to the observer in the new frame,
while when $\eta < 0$ the length appears contracted.

These transformation properties make sense when we recall the meaning behind
light front synchronization discussed in Sec.~\ref{sec:tilted}.
To an observer downstream of (with a lesser $\tilde{z}$ coordinate than) the system under observation,
$\eta > 0$ means the system is moving away from the observer (in the $+z$ direction),
and $\eta < 0$ means the system is moving towards them.
For $\eta > 0$, the dilation of time and length of a moving system corresponds to a redshift
while the time quickening and length contraction of a system moving with $\eta < 0$
corresponds to a blueshift.
These are exactly the usual relativistic redshift and blueshift.

The Poincar\'{e} group has long been known to have a $(2+1)$-dimensional Galilean
subgroup~\cite{Bargmann:1954gh,Pinski:1968galilean,Soper:1972xc,Takahashi:1988jz,Burkardt:2002hr}.
This subgroup is generated by the basis
$\{\tilde{E}, \tilde{P}^i_\perp, \tilde{P}_z, \tilde{J}_3, \tilde{B}_{\perp}^i\}$,
where the $\perp$ subscript means $i$ only runs over $1$ and $2$.
Under this group, the time $\tilde{\tau}$ remains invariant.
The algebra of this subgroup is:
\begin{subequations}
  \begin{align}
    [\tilde{J}_3, \tilde{B}^i_\perp]
    &=
    \i \epsilon_{ij3} \tilde{B}^j_\perp
    \\
    [\tilde{B}_\perp^i, \tilde{B}_\perp^j]
    &=
    0
    \\
    [\tilde{J}_3, \tilde{P}^i_\perp]
    &=
    \i \epsilon_{ij3} \tilde{P}^j_\perp
    \\
    [\tilde{B}^i_\perp, \tilde{P}^j_\perp]
    &=
    \i \delta_{ij} \tilde{P}_z
    \\
    [\tilde{B}^i_\perp, \tilde{E}]
    &=
    \i \tilde{P}^i_\perp
    \,,
  \end{align}
\end{subequations}
with all other commutators zero.
Notably, $\tilde{P}_z$ commutates with all other generators in the group.
It is accordingly considered a central charge of the group,
and this observation is identical to the well-known fact that $P^+$
is a central charge of the Galilean subgroup in light front coordinates~\cite{Soper:1972xc,Burkardt:2002hr}.
Physically, this means that the transformations in the Galilean subgroup
(including all translations, transverse boosts, and rotations around the $z$ axis)
will leave $\tilde{P}_z$ unaltered.

The Galilean subgroup has another central charge,
which is related to the Pauli-Lubanski axial four-vector~\cite{lubanski1942theorie1,lubanski1942theorie2}:
\begin{align}
  W_\mu
  =
  -\frac{1}{2}
  \epsilon_{\mu\nu\rho\sigma}
  J^{\nu\rho} P^\sigma
  \,,
\end{align}
defined with the convention that $\epsilon^{0123} = +1$.
For its expression in tilted coordinates,
one need only put tildes over each of the tensors involved.
Just as with momentum, we define individual components of the associated
three-vector using the covariant components of the Pauli-Lubanski vector:
$\tilde{W}_\mu = (\tilde{W}_0; -\tilde{W}_x,  -\tilde{W}_y,  -\tilde{W}_z)$.
Through some algebra, one can show that:
\begin{align}
  \tilde{W}_z
  =
  \tilde{P}_z \tilde{J}_3
  -
  (\tilde{\bm{B}}_\perp \times \tilde{\bm{P}}_\perp)\cdot\hat{z}
  \,,
\end{align}
and additionally one can show that $\tilde{W}_z$ commutes
with all the generators of the Galilean subgroup,
making it another central charge of this group.
Additionally,
\begin{align}
  \label{eqn:hel:op}
  \hat{\lambda}
  =
  \tilde{P}_z^{-1}
  \tilde{W}_z
  =
  \tilde{J}_3
  -
  \tilde{P}_z^{-1}
  (\tilde{\bm{B}}_\perp \times \tilde{\bm{P}}_\perp)\cdot\hat{z}
\end{align}
is invariant under the Galilean subgroup,
and is equal to the $z$ component of angular momentum for a system
at transverse rest.
The operator $\hat{\lambda}$ is the standard light front helicity operator~\cite{Soper:1972xc},
and characterizes the $z$ component of intrinsic spin in a Galilean-invariant manner.
In particular---as pointed out in Ref.~\cite{Soper:1972xc}---this
operator characterizes with spin projection along the $z$ axis
when the system is at rest.

One can show that, under longitudinal boosts:
\begin{subequations}
  \begin{align}
    [\tilde{B}_3, \tilde{P}_z]
    &=
    i \tilde{P}_z
    \\
    [\tilde{B}_3, \tilde{W}_z]
    &=
    i \tilde{W}_z
    \\
    [\tilde{B}_3, \hat{\lambda}]
    &=
    0
    \,.
  \end{align}
\end{subequations}
The light front helicity operator $\hat{\lambda}$ is invariant under longitudinal boosts.
This makes light front helicity an especially useful label for characterizing the spin states of hadrons:
a rest frame state with definite $\lambda$ can be transformed to an arbitrary momentum state with the same $\lambda$
by performing a longitudinal boost followed by a transverse boost.
We will therefore use light front helicity to label the spin content of hadrons
throughout the remainder of this work.


\subsection{Conserved currents and densities at fixed light front time}
\label{sec:cc}

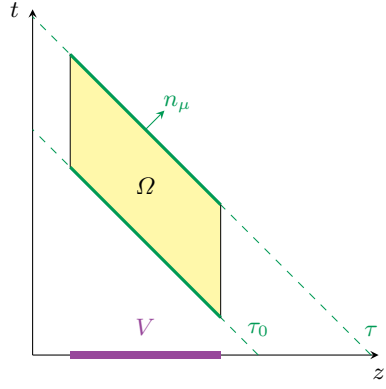
\begin{figure}
  \centering
  \begin{tikzpicture}
    \draw[-stealth] (0.5,-0.5) -- (0.5, 4.1) node[left=2pt]  {$t$} ;
    \draw[-stealth] (0.5,-0.5) -- (5.1, -0.5) node[below=2pt] {$z$};
    \draw[dashed, ForestGreen] (3.5, -0.5) node[above=4pt] {$\tau_0$} -- (0.5,2.5);
    \draw[dashed, ForestGreen] (5, -0.5) node[above=4pt] {$\tau$} -- (0.5,4);
    \draw[line width=0,fill=Yellow!42] (1,2) -- (3,0) -- (3,1.5) -- (1,3.5) -- cycle;
    \draw[very thick, ForestGreen] (1,2) -- (3,0);
    \draw[very thick, ForestGreen] (1,3.5) -- node[above right,anchor=south west](n){}(3,1.5);
    \node (O) at (2,1.75) {$\varOmega$};
    \path (n.south west) edge[-stealth,ForestGreen] node[above right] {$n_\mu$} (n.north east);
    \draw[very thick, Purple] (2,-0.5) node[above=4pt] {$V$} -- (1,-0.5);
    \fill[Purple] (3,-0.45) rectangle (1,-0.55);
  \end{tikzpicture}
  \caption{
    A finite spacetime region $\varOmega$ bounded by two hypersurfaces
    of equal light front time $\tau_0$ and $\tau$.
    Each slice of fixed light front time contains the same spatial region $V$.
    The future-directed normal $n_\mu$ to the equal-light-front-time hypersurfaces
    is also indicated in this diagram.
  }
  \label{fig:spacetime}
\end{figure}

Before proceeding to the construction of internal hadronic densities,
it is important to understand how conserved currents should be understood
in terms of light front time,
especially for correctly identifying which component should be identified as
the static density (e.g., as the charge density or as the energy density).
A conserved current obeys the differential continuity equation:
\begin{align}
  \partial_\mu j^\mu(x)
  =
  \tilde{\partial}_\mu \tilde{j}^\mu(x)
  =
  0
  \,,
\end{align}
which is the same in Minkowski and tilted coordinates.
We will first proceed using Minkowski coordinates.
The continuity equation can be cast into integral form by integrating
it over a spacetime region $\varOmega$:
\begin{align}
  \int_{\varOmega} \d^4x \,
  \partial_\mu j^\mu(x)
  =
  0
  \,.
\end{align}
The specific region of spacetime considered consists of a spatial volume
$V$ bounded by two equal-light-front-time hypersurfaces
at $x^+ = \tau_0$ and $x^+=\tau$,
and is depicted in Fig.~\ref{fig:spacetime}.
Using the divergence theorem,
the integral of the continuity equation over $\varOmega$ can be rewritten
as an integral over its surface.
If we take $n_\mu$ to be the future-pointing normal vector to the
equal-light-front-time surfaces, then:
\begin{align}
  \int_V \d^3\bm{x} \,
  n_\mu j^\mu(x)\bigg|_{x^+=\tau}
  -
  \int_V \d^3\bm{x} \,
  n_\mu j^\mu(x)\bigg|_{x^+=\tau_0}
  +
  \int_{\tau_0}^{\tau} \d\tau'
  \int_{\partial V} \d \bm{S} \cdot
  \bm{j}(x)\bigg|_{x^+=\tau'}
  =
  0
  \,,
\end{align}
and then differentiating this with respect to $\tau$ gives:
\begin{align}
  \frac{\d}{\d\tau}
  \left[
    \int_V \d^3\bm{x} \,
    n_\mu j^\mu(x)\bigg|_{x^+=\tau}
    \right]
  =
  -
  \int_{\partial V} \d \bm{S} \cdot
  \bm{j}(x)\bigg|_{x^+=\tau}
  \,.
\end{align}
This form has a clear interpretation as the integral form of a
continuity equation.
The left-hand side quantifies how the amount of ``charge''
in a spatial region $V$ is changing with time,
and the right-hand side quantifies how much current is flowing through
the boundary of this spatial region at a fixed light front time $\tau$.
The right-hand side becomes zero if $V$ is all of space
and $j^\mu$ is localized in space
(as is expected of physically realistic currents),
and in this case the equation is just a statement of charge conservation.
Since this relation must hold for any spatial region $V$,
the static charge density is given by
\begin{align}
  \rho(x)
  =
  n_\mu j^\mu(x)
  =
  j^+(x)
\end{align}
when expressed in Minkowski coordinates,
which is identical to the standard light front charge density.
This is the logical result of using light front time synchronization,
and not an arbitrary choice.

The same derivation can be done using tilted instead of Minkowski coordinates.
Fixed-time surfaces are given by fixed $\tilde{\tau}$,
and the future-pointing normal to equal light front time surfaces gives
\begin{align}
  \rho(x)
  =
  \tilde{n}_\mu \tilde{j}^\mu(x)
  =
  \tilde{j}^0(x)
  \,.
\end{align}
Comparing to Eq.~(\ref{eqn:tilted}),
one can see that $\tilde{j}^0(x) = j^+(x)$,
so these coordinate systems give consistent results
if the same equal-light-front-time surface is used in both.
The vector current is given by $\tilde{\bm{j}}(x) = \bm{j}(x)$
regardless of synchronization convention.


\section{Wave packets and probability currents in tilted coordinates}
\label{sec:packet}

To isolate purely internal properties of composite spin-half systems,
we require a formulation for the wave function describing the location
of the system itself.
We shall construct that formulation here,
along with four-currents for the probability and spin projections,
which will appear in Sec.~\ref{sec:theory} as smearing functions.

We assume that it is possible to prepare a state with a single
particle of the species under consideration;
we consider a proton for concreteness.
Single-proton states are characterized by
the three independent components of their four-momentum,
namely $\tilde{\bm{p}}$,
as well as their light front helicity $\lambda$.
As noted by BBM~\cite{Blunden:1999wb},
light front quantization can be used with tilted coordinates,
since equal-time surfaces are defined the same way in both systems.
The creation and annihilation operators for the proton field obey the
canonical anticommutation relations:
\begin{align}
  \{ b(\tilde{\bm{p}},\lambda), b^\dagger(\tilde{\bm{p}}',\lambda') \}
  =
  \{ d(\tilde{\bm{p}},\lambda), d^\dagger(\tilde{\bm{p}}',\lambda') \}
  =
  2\tilde{p}_z
  (2\pi)^3 \delta^{(3)}(\tilde{\bm{p}} - \tilde{\bm{p}}')
  \delta_{\lambda\lambda'}
  \,,
\end{align}
and all other anticommutators are zero.
Here, $b$ and $b^\dagger$ annihilate and create protons,
while $d$ and $d^\dagger$ annihilate and create anti-protons.
Single-proton states of definite momentum are given by:
\begin{align}
  |\tilde{\bm{p}},\lambda\rangle
  =
  b^\dagger(\tilde{\bm{p}},\lambda)
  |0\rangle
  \,,
\end{align}
and the projection operator for identifying single-proton states is:
\begin{align}
  \label{eqn:completeness}
  \varLambda_{\text{1p}}
  =
  \sum_\lambda
  \int \frac{\d^3\tilde{\bm{p}}}{2\tilde{p}_z(2\pi)^3}
  |\tilde{\bm{p}},\lambda \rangle
  \langle \tilde{\bm{p}},\lambda|
  \,.
\end{align}

Let $|\varPsi\rangle$ signify a normalized
single-proton state, meaning:
\begin{subequations}
  \begin{align}
    \langle\varPsi|\varPsi\rangle
    &=
    1
    \\
    |\varPsi\rangle
    &=
    \varLambda_{\text{1p}}
    |\varPsi\rangle
    \,.
  \end{align}
\end{subequations}
When dealing with such a state, we can treat $\varLambda_{\text{1p}}$ as an
identity operator and use Eq.~(\ref{eqn:completeness}) as a
completeness relation.
The momentum-representation wave function can be defined as
$\langle \tilde{\bm{p}}, \lambda | \varPsi \rangle$,
which obeys the normalization rule:
\begin{align}
  \label{eqn:norm:mom}
  \sum_\lambda
  \int \frac{\d^3\tilde{\bm{p}}}{2\tilde{p}_z(2\pi)^3}
  \big| \langle \tilde{\bm{p}}, \lambda | \varPsi \rangle \big|^2
  =
  1
  \,.
\end{align}

The normal mode expansion of the proton field operator is given by:
\begin{align}
  \hat{\psi}(x)
  =
  \sum_\lambda
  \int \frac{\d^3\tilde{\bm{p}}}{2\tilde{p}_z(2\pi)^3}
  \Big\{
    u(\tilde{\bm{p}},\lambda)
    \e^{-\i\tilde{p}\cdot\tilde{x}}
    b(\tilde{\bm{p}},\lambda)
    +
    v(\tilde{\bm{p}},\lambda)
    \e^{+\i\tilde{p}\cdot\tilde{x}}
    d^\dagger(\tilde{\bm{p}},\lambda)
    \Big\}
  \,,
\end{align}
where the spinors $u$ and $v$ are normalized such that
$\bar{u}(\tilde{\bm{p}},\lambda') u(\tilde{\bm{p}},\lambda) = 2m\delta_{\lambda\lambda'}$
and
$\bar{v}(\tilde{\bm{p}},\lambda') v(\tilde{\bm{p}},\lambda) = -2m\delta_{\lambda\lambda'}$\footnote{
  We remark for clarity that
  $\bar{u}(\tilde{\bm{p}},\lambda)
  =
  u^\dagger(\tilde{\bm{p}},\lambda) \gamma^0$
  and
  $\bar{v}(\tilde{\bm{p}},\lambda)
  =
  v^\dagger(\tilde{\bm{p}},\lambda) \gamma^0$
  as usual, and that $\tilde{\gamma}^0$ is not used to define the barred spinors.
  In the context of defining barred spinors,
  $\gamma^0$ plays the role of swapping left-handed and right-handed components,
  rather than as the $0$th component of a four-vector.
  Put another way,
  $\bar{u}(\tilde{\bm{p}}',\lambda')
  u(\tilde{\bm{p}},\lambda)$
  is a scalar, and therefore should not change with the transformation
  from Minkowski to tilted coordinates,
  meaning the definition of
  $\bar{u}(\tilde{\bm{p}},\lambda)$ should not change to use $\tilde{\gamma}^0$.
}.
Explicit expressions for the spinors can be found in Appendix~\ref{sec:spinors}.
A fully covariant coordinate-representation wave function for a single-proton state
can be defined through the field operator as:
\begin{align}
  \varPsi_{\text{cov}}(x)
  =
  \langle 0 | \hat{\psi}(x) | \varPsi \rangle
  \,,
\end{align}
and is related to the momentum-representation wave function
through a Fourier transform:
\begin{align}
  \varPsi_{\text{cov}}(x)
  =
  \sum_\lambda
  \int \frac{\d^3\tilde{\bm{p}}}{2\tilde{p}_z(2\pi)^3}
  u(\tilde{\bm{p}},\lambda)
  \langle \tilde{\bm{p}}, \lambda | \varPsi\rangle
  \e^{-\i\tilde{p}\cdot\tilde{x}}
  \,.
\end{align}
Through this relation, it can be shown that
$\varPsi(\tilde{\bm{x}},\tilde{\tau})$
obeys the Dirac equation:
\begin{align}
  (\i\slashed{\partial}-m)
  \varPsi_{\text{cov}}(x)
  =
  \sum_\lambda
  \int \frac{\d^3\tilde{\bm{p}}}{2\tilde{p}_z(2\pi)^3}
  (\slashed{p}-m)
  u(\tilde{\bm{p}},\lambda)
  \langle \tilde{\bm{p}}, \lambda | \varPsi\rangle
  \e^{-\i\tilde{p}\cdot\tilde{x}}
  =
  0
  \,,
\end{align}
where $(\slashed{p}-m)u(\tilde{\bm{p}},\lambda)=0$ was used.


\subsection{Helicity-component wave functions}

It is helpful to consider helicity components of the proton wave function.
These are defined:
\begin{align}
  \label{eqn:fourier}
  \varPsi_\lambda(\tilde{\bm{x}},\tilde{\tau})
  =
  \frac{ \bar{u}(\i\tilde{\bm{\nabla}},\lambda) }{2m}
  \varPsi_{\text{cov}}(x)
  =
  \int \frac{\d^3\tilde{\bm{p}}}{2\tilde{p}_z(2\pi)^3}
  \langle \tilde{\bm{p}}, \lambda | \varPsi\rangle
  \e^{-\i\tilde{p}\cdot\tilde{x}}
  \,,
\end{align}
and are given by a Fourier transform of the momentum-representation wave function.
From inverting Eq.~(\ref{eqn:fourier}) and placing it into
Eq.~(\ref{eqn:norm:mom}),
the following normalization relation can be derived for the helicity components:
\begin{align}
  \sum_\lambda
  \int \d^3 \tilde{\bm{x}} \,
  \varPsi^*_\lambda(\tilde{\bm{x}},\tilde{\tau})
  \i \overleftrightarrow{\tilde{\partial}}^0
  \varPsi_\lambda(\tilde{\bm{x}},\tilde{\tau})
  =
  1
  \,,
\end{align}
where $\tilde{\partial}^0 \equiv \tilde{g}^{0\mu} \tilde{\partial}_\mu$
and where the two-sided derivative is defined as
$f\overleftrightarrow{\partial}_\mu g
= f (\partial_\mu g) - (\partial_\mu f)g$.
Notably, this appears to be the $0$th component of a four-current:
\begin{align}
  \label{eqn:prob}
  \mathscr{P}^\mu(\tilde{\bm{x}},\tilde{\tau})
  =
  \sum_\lambda
  \varPsi^*_\lambda(\tilde{\bm{x}},\tilde{\tau})
  \i \overleftrightarrow{\tilde{\partial}}^\mu
  \varPsi_\lambda(\tilde{\bm{x}},\tilde{\tau})
  \,,
\end{align}
and it can be confirmed through the help of Eq.~(\ref{eqn:fourier}) that:
\begin{align}
  \tilde{\partial}_\mu
  \mathscr{P}^\mu(\tilde{\bm{x}},\tilde{\tau})
  =
  0
  \,.
\end{align}
Thus $\mathscr{P}^\mu(\tilde{\bm{x}},\tilde{\tau})$ is a conserved current.
As discussed in Sec.~\ref{sec:cc},
$\mathscr{P}^0(\tilde{\bm{x}},\tilde{\tau})$
is thus the density of the conserved ``charge'' associated with this current.
Its normalization to unity confers it an interpretation as a probability density,
and thus $\mathscr{P}^\mu(\tilde{\bm{x}},\tilde{\tau})$
can be interpreted as the probability four-current.


\subsection{Spin density matrices}
\label{sec:spin}

Lastly, we define spin-projection densities and currents.
In this formulation, we emphasize properties that are invariant under
light front boosts
and use linear combinations of definite-$\lambda$ states to
define spin densities.
In particular, we follow the conventional definitions
in the light front literature~\cite{Carlson:2007xd}:
\begin{subequations}
  \begin{align}
    |s_z = \pm 1/2\rangle
    &=
    |\lambda = \pm 1/2\rangle
    \\
    |s_x = \pm 1/2\rangle
    &=
    \frac{
      |\lambda = + 1/2\rangle
      \pm
      |\lambda = - 1/2\rangle
    }{\sqrt{2}}
    \\
    |s_y = \pm 1/2\rangle
    &=
    \frac{
      |\lambda = + 1/2\rangle
      \pm \i
      |\lambda = - 1/2\rangle
    }{\sqrt{2}}
    \,.
  \end{align}
\end{subequations}
It should be remarked that for systems with non-zero velocity,
these are not actually eigenstates of the appropriate components
of the Pauli-Lubanski vector.
However, for a system at rest,
these do become true spin eigenstates.
Moreover, these states are invariant under light front boosts,
and accordingly can be identified as states with
definite spin in their own rest frame.

States with definite spin along the $a$ axis
(where $a=x,y,z$) are eigenstates of the Pauli matrix $\sigma_a$,
and accordingly a spin density can be defined:
\begin{align}
  \label{eqn:spin}
  \mathscr{S}_a(\tilde{\bm{x}},\tilde{\tau})
  =
  \sum_{\lambda\lambda'}
  \varPsi^*_{\lambda'}(\tilde{\bm{x}},\tilde{\tau})
  \i \overleftrightarrow{\tilde{\partial}}^0
  (\sigma_a)_{\lambda'\lambda}
  \varPsi_\lambda(\tilde{\bm{x}},\tilde{\tau})
  \,.
\end{align}
This is essentially a density for the expectation value of $\sigma_a$.
Just as with the probability density,
this is the $0$th component of a four-current that is conserved.

Since spin, defined in this manner, is boost-invariant,
it is meaningful to consider it an intrinsic property of
the system under consideration.
It is thus helpful to consider a spin density matrix:
\begin{align}
  \label{eqn:density:matrix}
  \mathscr{P}_{\lambda'\lambda}^\mu(\tilde{\bm{x}},\tilde{\tau})
  =
  \varPsi^*_{\lambda'}(\tilde{\bm{x}},\tilde{\tau})
  \i \overleftrightarrow{\tilde{\partial}}^\mu
  \varPsi_\lambda(\tilde{\bm{x}},\tilde{\tau})
  \,,
\end{align}
the trace of which gives the probability four-current
of Eq.~(\ref{eqn:prob}).
Spin four-currents can similarly be obtained from multiplying
by $(\sigma_a)_{\lambda'\lambda}$ and summing over $\lambda,\lambda'$,
with the $0$th component of this operation reproducing
the spin-projection density in Eq.~(\ref{eqn:spin}).


\section{Derivation of electromagnetic densities}
\label{sec:theory}

In this section, we derive the internal electromagnetic densities of the proton
when light front time synchronization is used.
We closely follow the methodology of Ref.~\cite{Freese:2022fat}:
the physical density is identified as the expectation value of the
electromagnetic current operator $\hat{j}^\mu(x)$ within a physical state,
which is decomposed into a sum of wave-packet-independent internal densities
convolved with packet-dependent smearing functions.


\subsection{General formula for the physical electromagnetic current}
\label{sec:derive}

The physical electromagnetic current is the expectation value
of the electromagnetic current operator for a physical state:
\begin{align}
  \langle \tilde{j}^\mu(\tilde{x}) \rangle
  &=
  \langle \varPsi | \hat{j}^\mu(\tilde{x}) | \varPsi \rangle
  \,.
\end{align}
Using $\hat{j}^\mu(\tilde{x}) = \e^{\i\hat{P}\cdot\tilde{x}}\hat{j}^\mu(0)\e^{-\i\hat{P}\cdot\tilde{x}}$
and the one-particle projection operator, we can write:
\begin{align}
  \langle \tilde{j}^\mu(\tilde{x}) \rangle
  &=
  \sum_{\lambda,\lambda'}
  \int \frac{\d^3\tilde{\bm{p}}}{2\tilde{p}_z (2\pi)^3}
  \int \frac{\d^3\tilde{\bm{p}}'}{2\tilde{p}'_z (2\pi)^3}
  \varPsi^*(\tilde{\bm{p}}',\lambda')
  \langle \tilde{\bm{p}}', \lambda' | \hat{j}^\mu(0) | \tilde{\bm{p}}, \lambda \rangle
  \varPsi  (\tilde{\bm{p}} ,\lambda )
  \e^{\i\tilde{\varDelta} \cdot\tilde{x}}
  \,,
\end{align}
where $\tilde{\varDelta} = \tilde{p}' - \tilde{p}$.
Using the inversion of Eq.~(\ref{eqn:fourier}) here gives:
\begin{align}
  \langle \tilde{j}^\mu(\tilde{x}) \rangle
  &=
  \sum_{\lambda,\lambda'}
  \int \d^3\tilde{\bm{R}}
  \int \d^3\tilde{\bm{R}}'
  \int \frac{\d^3\tilde{\bm{P}}}{(2\pi)^3}
  \int \frac{\d^3\tilde{\bm{\varDelta}}}{(2\pi)^3}
  \varPsi_{\lambda'}^*(\tilde{\bm{R}}',\tilde{\tau})
  \langle \tilde{\bm{p}}', \lambda' | \hat{j}^\mu(0) | \tilde{\bm{p}}, \lambda \rangle
  \varPsi_\lambda  (\tilde{\bm{R}} , \tilde{\tau})
  \e^{-\i\tilde{\bm{\varDelta}} \cdot(\tilde{\bm{x}}-\frac{\tilde{\bm{R}}+\tilde{\bm{R}}'}{2})}
  \e^{+\i\tilde{\bm{P}} \cdot(\tilde{\bm{R}}'-\tilde{\bm{R}})}
  \,,
\end{align}
where $\tilde{P} = \frac{1}{2}\big(\tilde{p} + \tilde{p}'\big)$.
As observed in Refs.~\cite{Li:2022ldb,Freese:2022fat},
the integral over $\tilde{\bm{P}}$ can be performed, producing a delta function that
sets $\tilde{\bm{R}}' = \tilde{\bm{R}}$,
if the substitutions
\begin{align}
  \label{eqn:substitute}
  \tilde{P}
  \mapsto
  \frac{\i}{2} \overleftrightarrow{\tilde{\partial}}
\end{align}
are implicitly made within the matrix element
$\langle \tilde{\bm{p}}', \lambda' | \hat{j}^\mu(0) | \tilde{\bm{p}}, \lambda \rangle$.
The result is:
\begin{align}
  \langle \tilde{j}^\mu(\tilde{x}) \rangle
  &=
  \sum_{\lambda,\lambda'}
  \int \d^3\tilde{\bm{R}}
  \int \frac{\d^3\tilde{\bm{\varDelta}}}{(2\pi)^3}
  \varPsi_{\lambda'}^*(\tilde{\bm{R}},\tilde{\tau})
  \langle \tilde{\bm{p}}', \lambda' | \hat{j}^\mu(0) | \tilde{\bm{p}}, \lambda \rangle
  \varPsi_\lambda  (\tilde{\bm{R}} , \tilde{\tau})
  \e^{-\i\tilde{\bm{\varDelta}} \cdot(\tilde{\bm{x}}-\tilde{\bm{R}})}
  \,.
\end{align}
The matrix element appearing here can be decomposed into form factors as:
\begin{align}
  \langle \tilde{\bm{p}}', \lambda' | \hat{j}^\mu(0) | \tilde{\bm{p}}, \lambda \rangle
  =
  \bar{u}(\tilde{\bm{p}}',\lambda')
  \left\{
    \tilde{\gamma}^\mu
    F_1(\varDelta^2)
    +
    \frac{\i \tilde{\sigma}^{\mu\nu} \tilde{\varDelta}_\nu}{2m}
    F_2(\varDelta^2)
    \right\}
  u(\tilde{\bm{p}},\lambda)
  \,,
\end{align}
where we use the normalization convention that
$F_2(0) = \kappa$ gives the anomalous magnetic moment.
The Lorentz-invariant momentum transfer in tilted coordinates is:
\begin{align}
  \label{eqn:Delta2}
  \varDelta^2
  =
  2 \tilde{\varDelta}_z \tilde{\varDelta}_0
  -
  \tilde{\bm{\varDelta}}^2
  =
  2 \tilde{\varDelta}_z
  \big( \tilde{p}_0' - \tilde{p}_0 \big)
  -
  \tilde{\bm{\varDelta}}^2
  =
  2 \tilde{\varDelta}_z
  \left(
  \frac{m^2 + \left(\tilde{\bm{P}}+\frac{1}{2}\tilde{\bm{\varDelta}}\right)^2}{2\tilde{P}_z + \tilde{\varDelta}_z}
  -
  \frac{m^2 + \left(\tilde{\bm{P}}-\frac{1}{2}\tilde{\bm{\varDelta}}\right)^2}{2\tilde{P}_z - \tilde{\varDelta}_z}
  \right)
  -
  \tilde{\bm{\varDelta}}^2
  \,,
\end{align}
which clearly depends on $\tilde{\bm{P}}$.
This $\tilde{\bm{P}}$ dependence causes the argument of the form factors
to depend on the wave packet,
which provides an obstacle to separating internal structure from
wave packet dependence.
As an alternative to Eq.~(\ref{eqn:Delta2}),
one can transform $\varDelta^2$ into a d'Alembertian acting on the
operator as a whole (as is explored in Appendix~\ref{sec:unique}),
but the presence of a time derivative in the d'Alembertian
still impedes eliminating wave packet dependence from the form factor.
(See Appendix~\ref{sec:unique} for details on this approach.)

As previously observed in Ref.~\cite{Freese:2022fat},
this undesired $\tilde{\bm{P}}$ dependence can be eliminated by integrating out
$\tilde{z}$ in the physical density.
Doing so turns $\e^{-\i\tilde{\bm{\varDelta}}_z\tilde{z}}$ into a
Dirac delta function that sets $\tilde{\varDelta}_z = 0$.
We are left with a two-dimensional density:
\begin{align}
  \langle \tilde{j}^\mu(\tilde{\bm{x}}_\perp,\tilde{\tau}) \rangle_{\text{2D}}
  \equiv
  \int \d\tilde{z} \,
  \langle \tilde{j}^\mu(\tilde{x}) \rangle
  =
  \sum_{\lambda,\lambda'}
  \int \d^3\tilde{\bm{R}}
  \int \frac{\d^2\tilde{\bm{\varDelta}}_\perp}{(2\pi)^2}
  \varPsi_{\lambda'}^*(\tilde{\bm{R}},\tilde{\tau})
  \langle \tilde{\bm{p}}', \lambda' | \hat{j}^\mu(0) | \tilde{\bm{p}}, \lambda \rangle
  \varPsi_\lambda  (\tilde{\bm{R}} , \tilde{\tau})
  \e^{-\i\tilde{\bm{\varDelta}}_\perp \cdot(\tilde{\bm{x}}-\tilde{\bm{R}})}
  \,,
\end{align}
where the substitution of Eq.~(\ref{eqn:substitute}) is still
implicitly being applied.
With $\tilde{\bm{\varDelta}}_z = 0$, the Lorentz-invariant momentum transfer is just:
\begin{align}
  \varDelta^2 = -\tilde{\bm{\varDelta}}_\perp^2
  \,.
\end{align}

At this point, we can use the explicit spinor elements found in
Appendix~\ref{sec:spinors} to write:
\begin{multline}
  \label{eqn:breakdown}
  \langle \tilde{\bm{p}}', \lambda' | \hat{j}^\mu(0) | \tilde{\bm{p}}, \lambda \rangle
  =
  2 \tilde{P}^\mu
  \left(
  (\sigma_0)_{\lambda'\lambda}
  F_1(-\tilde{\bm{\varDelta}}_\perp^2)
  -
  \frac{
    \i
    \epsilon^{\alpha\beta\gamma\delta}
    \tilde{n}_\alpha \tilde{\bar{n}}_\beta
    \tilde{\varDelta}_\gamma (\sigma_\delta)_{\lambda'\lambda}
  }{2m}
  F_2(-\tilde{\bm{\varDelta}}_\perp^2)
  \right)
  \\
  -
  \frac{
    \i
    \epsilon^{\mu\nu\rho\sigma}
    \tilde{n}_\nu
    \tilde{P}_\rho \tilde{\varDelta}_\sigma
  }{(\tilde{P}\cdot\tilde{n})}
  (\sigma_3)_{\lambda'\lambda}
  G_M(-\tilde{\bm{\varDelta}}_\perp^2)
  +
  \frac{m\tilde{n}^\mu}{(\tilde{P}\cdot\tilde{n})}
  \i
  \epsilon^{\alpha\beta\gamma\delta}
  \tilde{n}_\alpha \tilde{\bar{n}}_\beta
  \tilde{\varDelta}_\gamma (\sigma_\delta)_{\lambda'\lambda}
  G_M(-\tilde{\bm{\varDelta}}_\perp^2)
  \,,
\end{multline}
where
$G_M(-\tilde{\bm{\varDelta}}_\perp^2) = F_1(-\tilde{\bm{\varDelta}}_\perp^2) + F_2(-\tilde{\bm{\varDelta}}_\perp^2)$
is the Sachs magnetic form factor,
and where $\tilde{n}_\mu = (1;0,0,0)$
and $\tilde{\bar{n}}^\mu = (1;0,0,0)$ are four-vectors
that project out the time component of $\tilde{x}^\mu$
and the energy component of $\tilde{p}_\mu$ respectively---and
in this sense are analogous to the $n$ and $\bar{n}$ vectors
used in standard light front coordinates.
Noting the substitution rule of Eq.~(\ref{eqn:substitute}),
wave packet dependence is indicated by $\tilde{P}$ dependence.
The wave packet dependence in the first line of Eq.~(\ref{eqn:breakdown})
can be identified as the spin density matrix defined in
Eq.~(\ref{eqn:density:matrix}).
For the second line, additional smearing functions must be defined:
\begin{subequations}
  \label{eqn:additional}
  \begin{align}
    \mathscr{J}_{\nu,\lambda'\lambda}(\tilde{\bm{R}},\tilde{\tau})
    &=
    2m
    \varPsi_{\lambda'}^*(\tilde{\bm{R}},\tilde{\tau})
    \frac{-\tilde{P}_\nu}{(\tilde{P}\cdot\tilde{n})}
    \varPsi_{\lambda}  (\tilde{\bm{R}} , \tilde{\tau})
    \\
    \mathscr{K}_{\lambda'\lambda}(\tilde{\bm{R}},\tilde{\tau})
    &=
    2m
    \varPsi_{\lambda'}^*(\tilde{\bm{R}},\tilde{\tau})
    \frac{m}{(\tilde{P}\cdot\tilde{n})}
    \varPsi_{\lambda}  (\tilde{\bm{R}} , \tilde{\tau})
    \,,
  \end{align}
\end{subequations}
where factors of $2m$ were introduced so that these functions
have the same units as the spin density matrix.
If we additionally define the three following internal densities:
\begin{subequations}
  \label{eqn:densities:em}
  \begin{align}
    \rho_{\lambda'\lambda}(\tilde{\bm{b}}_\perp)
    &=
    \int \frac{\d^2\tilde{\bm{\varDelta}}_\perp}{(2\pi)^2}
    \left(
    (\sigma_0)_{\lambda'\lambda}
    F_1(-\tilde{\bm{\varDelta}}_\perp^2)
    -
    \frac{
      \i
      \epsilon^{\alpha\beta\gamma\delta}
      \tilde{n}_\alpha \tilde{\bar{n}}_\beta
      \tilde{\varDelta}_\gamma (\sigma_\delta)_{\lambda'\lambda}
    }{2m}
    F_2(-\tilde{\bm{\varDelta}}_\perp^2)
    \right)
    \e^{-\i\tilde{\bm{\varDelta}}_\perp \cdot \tilde{\bm{b}}_\perp}
    \\
    j_{\perp,\lambda'\lambda}^{\mu\nu}(\tilde{\bm{b}}_\perp)
    &=
    -
    \int \frac{\d^2\tilde{\bm{\varDelta}}_\perp}{(2\pi)^2}
    \frac{
      \i
      \epsilon^{\mu\nu\rho\sigma}
      \tilde{n}_\rho
      \tilde{\varDelta}_\sigma
    }{2m}
    (\sigma_3)_{\lambda'\lambda}
    G_M(-\tilde{\bm{\varDelta}}_\perp^2)
    \e^{-\i\tilde{\bm{\varDelta}}_\perp \cdot \tilde{\bm{b}}_\perp}
    \\
    j_{\parallel,\lambda'\lambda}^{\mu}(\tilde{\bm{b}}_\perp)
    &=
    \int \frac{\d^2\tilde{\bm{\varDelta}}_\perp}{(2\pi)^2}
    \frac{
      \i
      \epsilon^{\alpha\beta\gamma\delta}
      \tilde{n}_\alpha \tilde{\bar{n}}_\beta
      \tilde{\varDelta}_\gamma (\sigma_\delta)_{\lambda'\lambda}
    }{2m}
    \tilde{n}^\mu
    G_M(-\tilde{\bm{\varDelta}}_\perp^2)
    \e^{-\i\tilde{\bm{\varDelta}}_\perp \cdot \tilde{\bm{b}}_\perp}
    \,,
  \end{align}
\end{subequations}
then the physical density can be written as:
\begin{multline}
  \label{eqn:current:general}
  \langle \tilde{j}^\mu(\tilde{\bm{x}}_\perp,\tilde{\tau}) \rangle_{\text{2D}}
  =
  \sum_{\lambda,\lambda'}
  \int \d^3\tilde{\bm{R}} \,
  \Bigg\{
    \mathscr{P}_{\lambda'\lambda}^\mu(\tilde{\bm{R}},\tilde{\tau})
    \rho_{\lambda'\lambda}(\tilde{\bm{x}}_\perp-\tilde{\bm{R}}_\perp)
    \\
    +
    \mathscr{J}_{\nu,\lambda'\lambda}(\tilde{\bm{R}},\tilde{\tau})
    j_{\perp,\lambda'\lambda}^{\mu\nu}(\tilde{\bm{x}}_\perp-\tilde{\bm{R}}_\perp)
    +
    \mathscr{K}_{\lambda'\lambda}(\tilde{\bm{R}},\tilde{\tau})
    j_{\parallel,\lambda'\lambda}^{\mu}(\tilde{\bm{x}}_\perp-\tilde{\bm{R}}_\perp)
    \Bigg\}
  \,.
\end{multline}
With the general expression now in hand,
we must now investigate the meaning of the internal densities
in Eq.~(\ref{eqn:densities:em}).


\subsection{Interpretation of internal densities}

Let us now consider the interpretation of the internal
densities in Eq.~(\ref{eqn:densities:em}),
with a special focus on the spatial structure of the proton
in its rest frame.
Although there is no rest state of the proton in Hilbert space---since
wave functions must necessarily be square integrable---we will show that
the internal densities are boost-invariant quantities,
and that artifacts of the proton's motion have all been absorbed into
the smearing functions in Eq.~(\ref{eqn:current:general}).
Moreover, these interpretations can be further validated
by utilizing the rest condition $\bm{P} = (0,0,m)$ found in Eq.~(\ref{eqn:rest}).

We shall consider states with definite light front spin $\frac{1}{2}\hat{\bm{s}}$---as
described in Sec.~\ref{sec:spin}---since
these correspond to a proton with definite spin in its own rest frame.
For these states, one can substitute:
\begin{subequations}
  \begin{align}
    (\sigma_0)_{\lambda'\lambda}
    &\mapsto
    1
    \\
    (\sigma_i)_{\lambda'\lambda}
    &\mapsto
    \hat{s}_i
  \end{align}
\end{subequations}
in the formulas of Eqs.~(\ref{eqn:density:matrix}),
(\ref{eqn:additional}) and (\ref{eqn:densities:em})
to obtain spin-state densities,
which will now be functions of $\hat{\bm{s}}$ instead of $\lambda$ and $\lambda'$.


\subsubsection{Internal charge density}

We consider the charge density first.
As discussed in Sec.~\ref{sec:cc},
the charge density can be identified by contracting the electromagnetic
current with $\tilde{n}_\mu$, giving the $0$th component.
Since
$j_{\perp}^{0\nu}(\tilde{\bm{b}}_\perp,\hat{\bm{s}}) = 0$
and
$j_{\parallel}^{0}(\tilde{\bm{b}}_\perp,\hat{\bm{s}}) = 0$
are both zero---a result that is frame-independent,
since boosts do not mix spatial components into light front time
(see Sec.~\ref{sec:lorentz})---the internal charge density is given
entirely by:
\begin{align}
  \label{eqn:rho}
  \rho(\tilde{\bm{b}}_\perp,\hat{\bm{s}})
  &=
  \int \frac{\d^2\tilde{\bm{\varDelta}}_\perp}{(2\pi)^2}
  \left(
  F_1(-\tilde{\bm{\varDelta}}_\perp^2)
  +
  \frac{
    (\hat{\bm{s}} \times \i\tilde{\bm{\Delta}}_\perp)\cdot\hat{z}
  }{2m}
  F_2(-\tilde{\bm{\varDelta}}_\perp^2)
  \right)
  \e^{-\i\tilde{\bm{\varDelta}}_\perp \cdot \tilde{\bm{b}}_\perp}
  \,,
\end{align}
where Eq.~(\ref{eqn:helpful}) was used to rewrite the terms multiplying
$F_2(-\tilde{\bm{\varDelta}}_\perp^2)$.
Notably, this reproduces the results of
Refs.~\cite{Burkardt:2002hr,Miller:2007uy,Miller:2009sg}
for longitudinally-polarized nucleons,
and (up to a sign discrepancy) the results of
Carlson and Vanderhaegen~\cite{Carlson:2007xd}
for transversely-polarized nucleons\footnote{
  The sign discrepancy is due to a mistake
  in Eq.~(1) of Ref.~\cite{Carlson:2007xd},
  where the complex exponential should be $\e^{-\i\bm{q}_\perp\cdot\bm{b}_\perp}$
  instead of $\e^{\i\bm{q}_\perp\cdot\bm{b}_\perp}$.
  The sign mistake originated in Ref.~\cite{Miller:2007uy},
  where only helicity states were considered so that this sign flip
  had no effect on its results.
}.
This internal charge density is smeared over space by the
probability current, as expected and as previously observed
for spin zero targets~\cite{Freese:2022fat}.
This probability current transforms as a Lorentz four-vector,
and effectively absorbs all of the boost dependence of the charge density
contribution to the physical electromagnetic four-current.
Since the spin degrees of freedom are characterized by light front helicity,
and since $\tilde{\bm{b}}_\perp = \tilde{\bm{x}}_\perp - \tilde{\bm{R}}_\perp$
is given by a difference between transverse spatial coordinates,
the internal density
$\rho(\tilde{\bm{b}}_\perp,\hat{\bm{s}})$
is fully invariant under both transverse and longitudinal boosts,
and therefore characterizes the internal charge density of the proton
in every reference frame---including its rest frame.


\subsubsection{Transverse current density}

We next consider the density
$j_{\perp}^{\mu\nu}(\tilde{\bm{b}}_\perp,\hat{\bm{s}})$.
Like the internal charge density,
it is boost-invariant,
with boost dependence having been absorbed into the smearing function
$\mathscr{J}_\nu(\tilde{\bm{R}},\tilde{\tau},\hat{\bm{s}})$.
The presence of $\tilde{n}_\rho$ in its definition ensures that the
$\mu=0$ and $\nu=0$ components are all zero.
The index $\mu$ signifies which component of the physical four-current
this density contributes to,
and $\nu$ signifies which component of the total four-momentum
this density is to be contracted with.
With the help of Eq.~(\ref{eqn:helpful}),
we can write out the non-zero components as:
\begin{subequations}
  \begin{align}
    j_{\perp}^{23}(\tilde{\bm{b}}_\perp,\hat{\bm{s}})
    =
    -
    j_{\perp}^{32}(\tilde{\bm{b}}_\perp,\hat{\bm{s}})
    &=
    +
    \hat{\bm{s}}\cdot\hat{z}
    \int \frac{\d^2\tilde{\bm{\varDelta}}_\perp}{(2\pi)^2}
    \frac{
      \i
      \tilde{\varDelta}_x
    }{2m}
    G_M(-\tilde{\bm{\varDelta}}_\perp^2)
    \e^{-\i\tilde{\bm{\varDelta}}_\perp \cdot \tilde{\bm{b}}_\perp}
    \\
    j_{\perp}^{13}(\tilde{\bm{b}}_\perp,\hat{\bm{s}})
    =
    -
    j_{\perp}^{31}(\tilde{\bm{b}}_\perp,\hat{\bm{s}})
    &=
    -
    \hat{\bm{s}}\cdot\hat{z}
    \int \frac{\d^2\tilde{\bm{\varDelta}}_\perp}{(2\pi)^2}
    \frac{
      \i
      \tilde{\varDelta}_y
    }{2m}
    G_M(-\tilde{\bm{\varDelta}}_\perp^2)
    \e^{-\i\tilde{\bm{\varDelta}}_\perp \cdot \tilde{\bm{b}}_\perp}
    \,.
  \end{align}
\end{subequations}
These entail, first of all, of a transverse current density:
\begin{align}
  \label{eqn:current:transverse}
  \bm{j}_{\perp}(\tilde{\bm{b}}_\perp,\hat{\bm{s}})
  =
  (\hat{\bm{s}}\cdot\hat{z})
  \int \frac{\d^2\tilde{\bm{\varDelta}}_\perp}{(2\pi)^2}
  \frac{
    (\hat{z} \times \i\tilde{\bm{\varDelta}}_\perp)
  }{2m}
  G_M(-\tilde{\bm{\varDelta}}_\perp^2)
  \e^{-\i\tilde{\bm{\varDelta}}_\perp \cdot \tilde{\bm{b}}_\perp}
  \,,
\end{align}
which, considering Eq.~(\ref{eqn:additional}), is smeared over
the transverse plane by the square of the wave function.
This smearing function is boost-invariant,
meaning the contribution of
$\bm{j}_{\perp}(\tilde{\bm{b}}_\perp,\hat{\bm{s}})$ to the physical
transverse current density is frame-independent.
This density can therefore be identified as the intrinsic transverse current density
of the proton, applicable to all frames including the rest frame of the proton.

In addition, the $\mu=3$ components of
$j_{\perp}^{\mu\nu}(\tilde{\bm{b}}_\perp,\hat{\bm{s}})$
entail a longitudinal current density:
\begin{align}
  j^3_{\text{ind.},i}(\tilde{\bm{b}}_\perp,\hat{\bm{s}})
  =
  (\hat{\bm{s}}\cdot\hat{z})
  \int \frac{\d^2\tilde{\bm{\varDelta}}_\perp}{(2\pi)^2}
  \frac{
    (\i\tilde{\bm{\varDelta}}_\perp \times \hat{z})_i
  }{2m}
  G_M(-\tilde{\bm{\varDelta}}_\perp^2)
  \e^{-\i\tilde{\bm{\varDelta}}_\perp \cdot \tilde{\bm{b}}_\perp}
  \,,
\end{align}
which---considering Eq.~(\ref{eqn:additional})---must be contracted with the expectation value of
the transverse velocity $(\tilde{\bm{P}}_{\perp} / \tilde{P}_z)_i$ to give a contribution to the physical
longitudinal current.
The smearing function associated with this contribution is not boost invariant,
and in fact vanishes in a frame where the expectation value of the transverse velocity is zero.
In fact, an intrinsic transverse current can induce a longitudinal current
under transverse boosts,
since transverse boosts mix transverse spatial coordinates into the $z$ coordinate---see
Eq.~(\ref{eqn:boost:transverse}).
Since
$j^3_{\text{ind.},i}(\tilde{\bm{b}}_\perp,\hat{\bm{s}})$
vanishes in a frame where the expected value of the transverse velocity is zero,
we call it a boost-induced longitudinal current,
in contrast to an intrinsic longitudinal current.
We shall address the intrinsic longitudinal current density next.


\subsubsection{Longitudinal current density}

The last intrinsic density defined in Eq.~(\ref{eqn:densities:em}) is
$j^\mu_\parallel$.
Considering $\tilde{n}^\mu = (0;0,0,-1)$,
the only component is $\mu=3$, which for definite-spin states can be written
(with the help of Eq.~(\ref{eqn:helpful})):
\begin{align}
  \label{eqn:current:longitudinal}
  j^3_\parallel(\tilde{\bm{b}}_\perp,\hat{\bm{s}})
  =
  \int \frac{\d^2\tilde{\bm{\varDelta}}_\perp}{(2\pi)^2}
  \frac{
    (\hat{\bm{s}}\times\i\tilde{\bm{\varDelta}}_\perp)\cdot\hat{z}
  }{2m}
  G_M(-\tilde{\bm{\varDelta}}_\perp^2)
  \e^{-\i\tilde{\bm{\varDelta}}_\perp \cdot \tilde{\bm{b}}_\perp}
  \,.
\end{align}
This is smeared over the transverse plane by the function
$\mathscr{H}(\tilde{\bm{R}},\tilde{\tau},\hat{\bm{s}})$
defined in Eq.~(\ref{eqn:additional}),
which is invariant under the Galilean subgroup (including transverse boosts)
but not under longitudinal boosts.
Under an \emph{active} boost of rapidity $\eta$,
the smearing function scales as $\e^{-\eta}$,
meaning the longitudinal current is redshifted (slowed down) for $\eta > 0$
and blueshifted (sped up) for $\eta < 0$.
We remind the reader than $\eta > 0$ increases $\tilde{P}_z$ and has the
system move away from the observer,
while $\eta < 0$ decreases $\tilde{P}_z$ and has the system move towards the
observer, assuming it is located upstream from
(at a larger $\tilde{z}$ coordinate than)
the observer.

Although the intrinsic transverse current and intrinsic longitudinal current
are separated in our formulation---the separation being necessary by virtue
of the different transformation properties of their smearing functions
under boosts---we may choose to formally combine
Eqs.~(\ref{eqn:current:transverse}) and (\ref{eqn:current:longitudinal})
into a single expression for the intrinsic electric current:
\begin{align}
  \label{eqn:current:internal}
  \tilde{\bm{j}}(\tilde{\bm{b}}_\perp,\hat{\bm{s}})
  =
  \int \frac{\d^2\tilde{\bm{\varDelta}}_\perp}{(2\pi)^2}
  \frac{ \hat{\bm{s}} \times \i \tilde{\bm{\varDelta}}_\perp }{2m}
  G_M(-\tilde{\bm{\varDelta}}_\perp^2)
  \e^{-\i\tilde{\bm{\varDelta}}_\perp \cdot \tilde{\bm{b}}_\perp}
  \,.
\end{align}


\section{Empirical rest frame charge and current densities}
\label{sec:empirical}

Using the results from Sec.~\ref{sec:theory},
we proceed to present numerical results for the internal
rest frame densities of the proton and the neutron.
For $F_1(\varDelta^2)$ and $F_2(\varDelta^2)$,
we use the parametrizations fit to empirical data by
Kelly~\cite{Kelly:2004hm},
combined with improved measurements of the neutron
electric form factor by Riordan \textsl{et al.}~\cite{Riordan:2010id}.

\begin{figure}
  \includegraphics[width=0.49\textwidth]{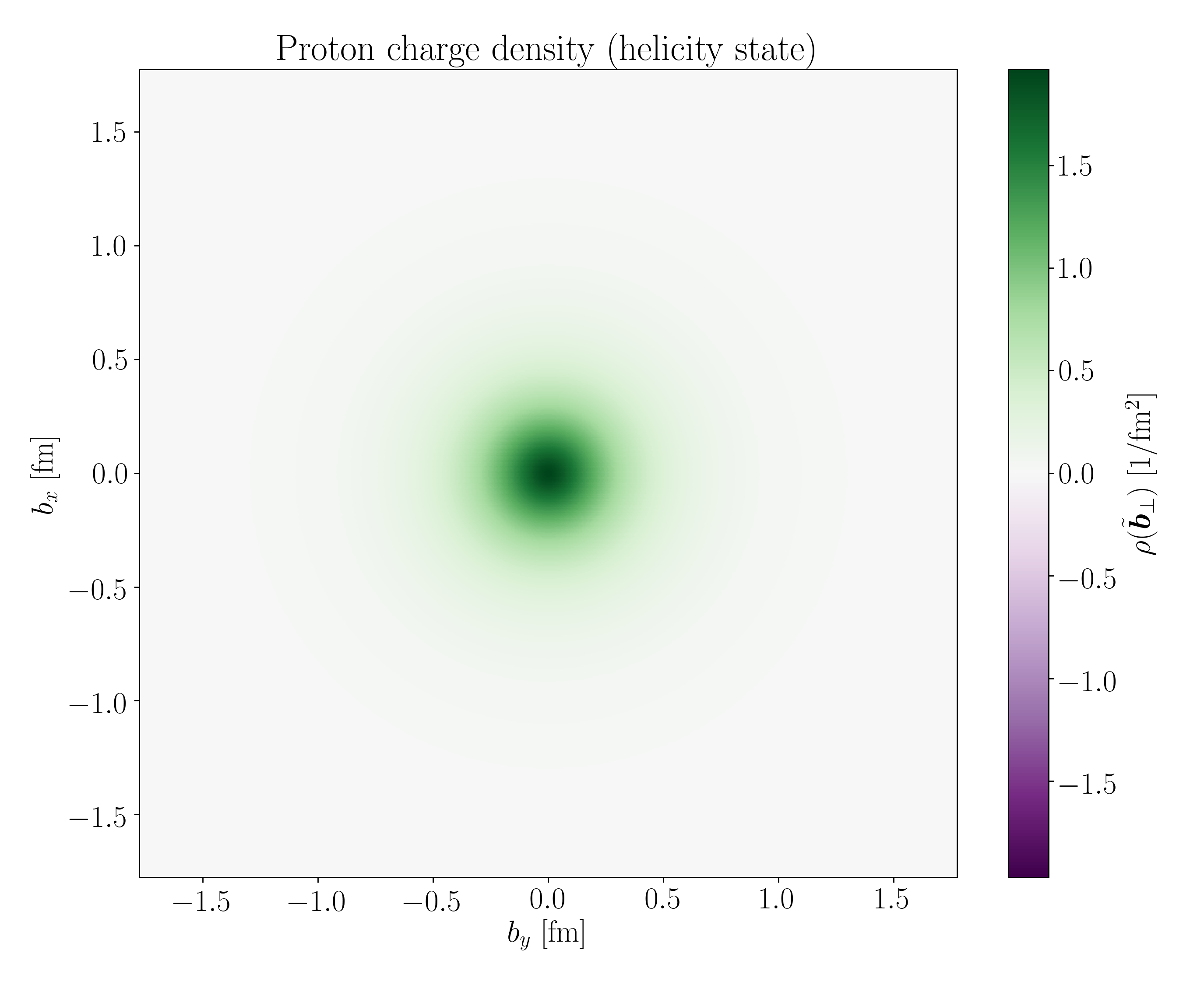}
  \includegraphics[width=0.49\textwidth]{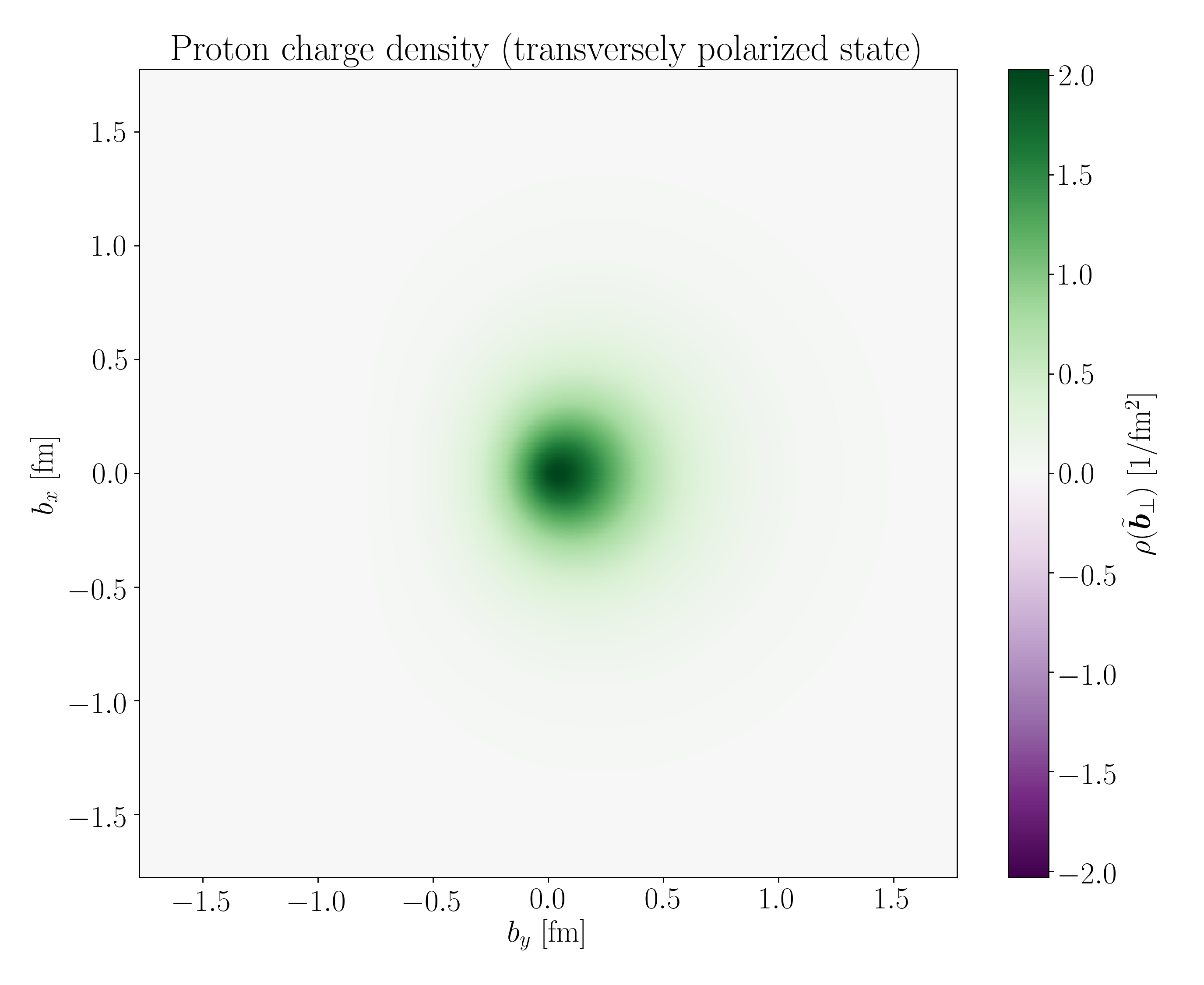}
  \includegraphics[width=0.49\textwidth]{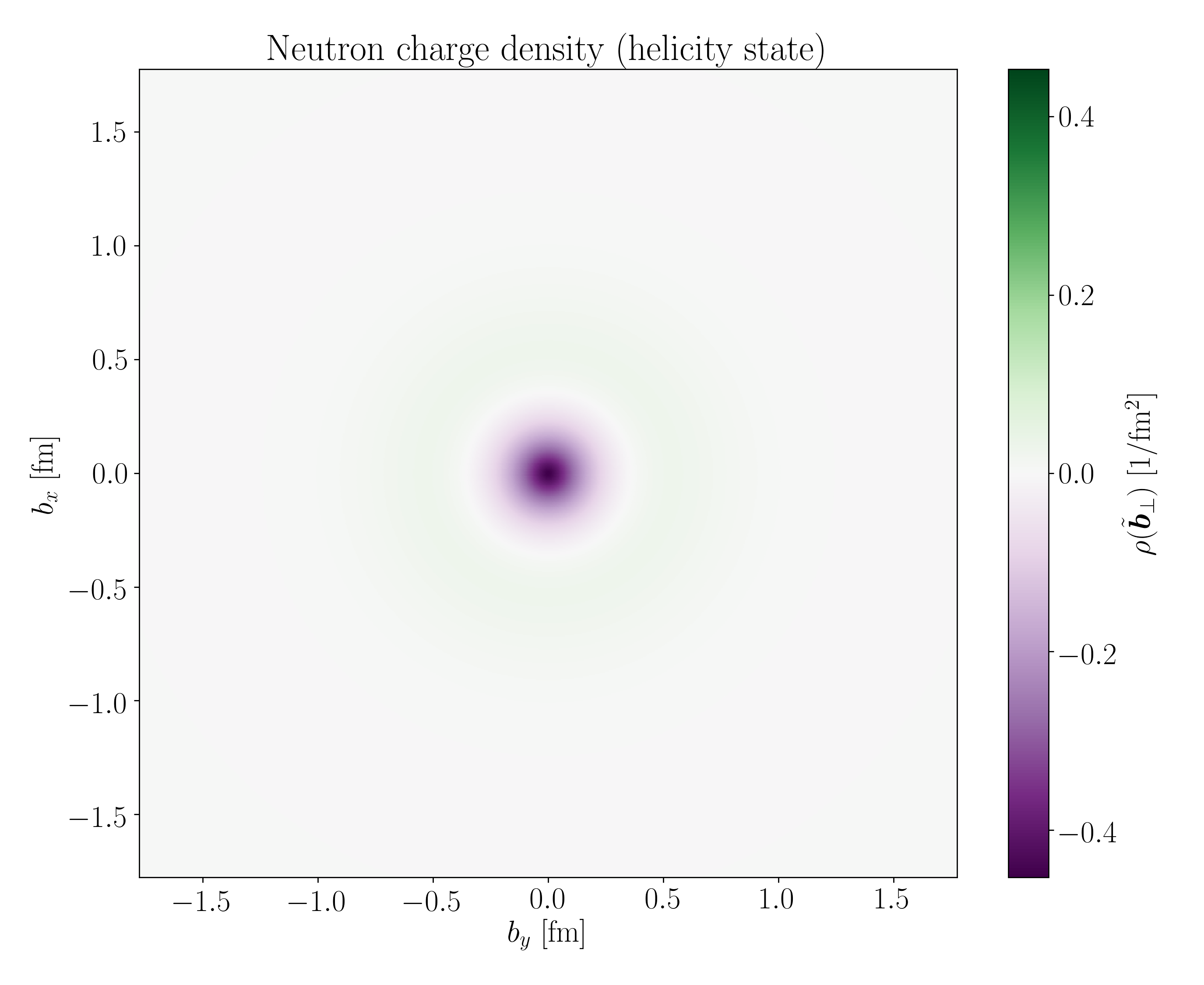}
  \includegraphics[width=0.49\textwidth]{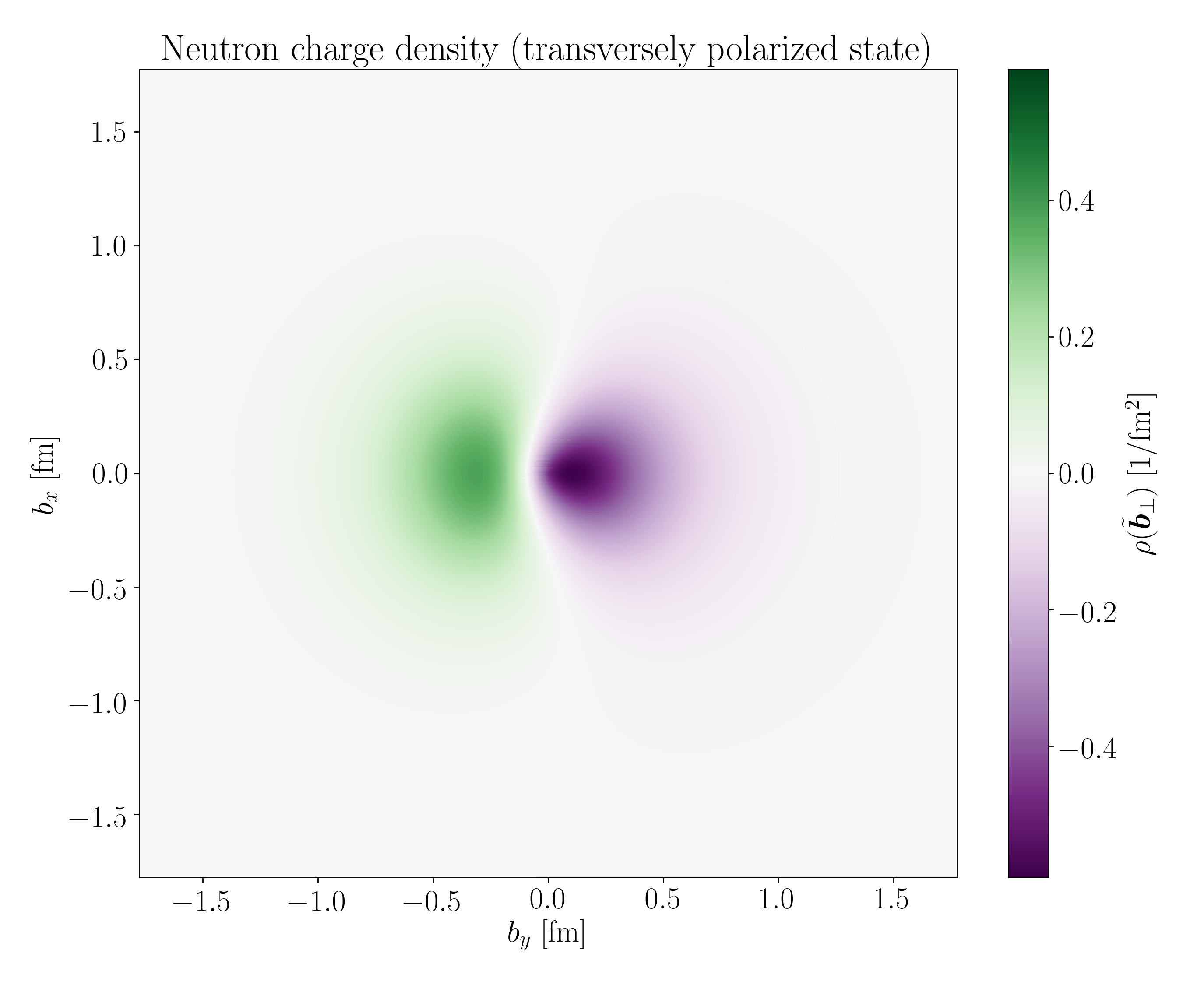}
  \caption{
    Rest frame charge densities of the proton (top row) and neutron (bottom row).
    The left column depicts positive-helicity states
    (spin-up along the $z$ axis),
    while the right column depicts transversely-polarized states
    with spin up along the $x$ axis.
  }
  \label{fig:charge}
\end{figure}

We first present results for the rest frame charge densities
in Fig.~\ref{fig:charge},
which can all be calculated using Eq.~(\ref{eqn:rho}).
The neutron charge density for helicity states reproduces the finding
of Miller~\cite{Miller:2007uy},
in which the neutron has a negatively-charged core surrounded by a diffuse
cloud of positive charge.
The charge densities for transversely-polarized states
reproduce the previous results
of Carlson and Vanderhaegen~\cite{Carlson:2007xd}
up to the previously noted sign discrepancy
in the $\sin\phi$ modulations.

It is worth emphasizing that we have reproduced and affirmed previously-known
results for the transverse charge density in the standard light front formalism.
These previous results were found to be entirely frame-independent,
with the spin degree of freedom specifically quantifying the nucleon's spin
in its own rest frame.
The limit of infinite momentum was never considered or taken.
It is mistaken to attribute features of these densities to kinematic effects
from boosting to infinite momentum.
The cause of angular modulations in the densities of transversely-polarized states
must lie elsewhere, and in fact originates from the
time synchronization convention.
We shall comment on this further in Sec.~\ref{sec:discuss}.

For now, we further remark that
the angular modulations in the charge densities of transversely-polarized states
produce an effective electric dipole moment,
which can be calculated from Eq.~(\ref{eqn:rho}) as follows:
\begin{align}
  \label{eqn:dipole}
  \bm{p}_{\text{eff}}
  \equiv
  \int \d^2\tilde{\bm{b}}_\perp \,
  \tilde{\bm{b}}_\perp
  \rho_{\text{rest}}(\tilde{\bm{b}}_\perp,\hat{\bm{s}})
  =
  \frac{\hat{z}\times\hat{\bm{s}}}{2m}
  F_2(0)
  =
  (\hat{z}\times\hat{\bm{s}})
  \frac{\kappa}{2m}
  \,,
\end{align}
where $\kappa$ is the anomalous magnetic moment of the nucleon.
Since it is an effect of using light front synchronization,
we call this a synchronization-induced electric dipole moment.
From the empirical values of the masses and anomalous magnetic moments,
the induced dipole moments of the proton and neutron are:
\begin{subequations}
  \begin{align}
    p_p
    &=
    0.19~\text{fm}
    \\
    p_n
    &=
    -0.20~\text{fm}
    \,.
  \end{align}
\end{subequations}

\begin{figure}
  \includegraphics[width=0.49\textwidth]{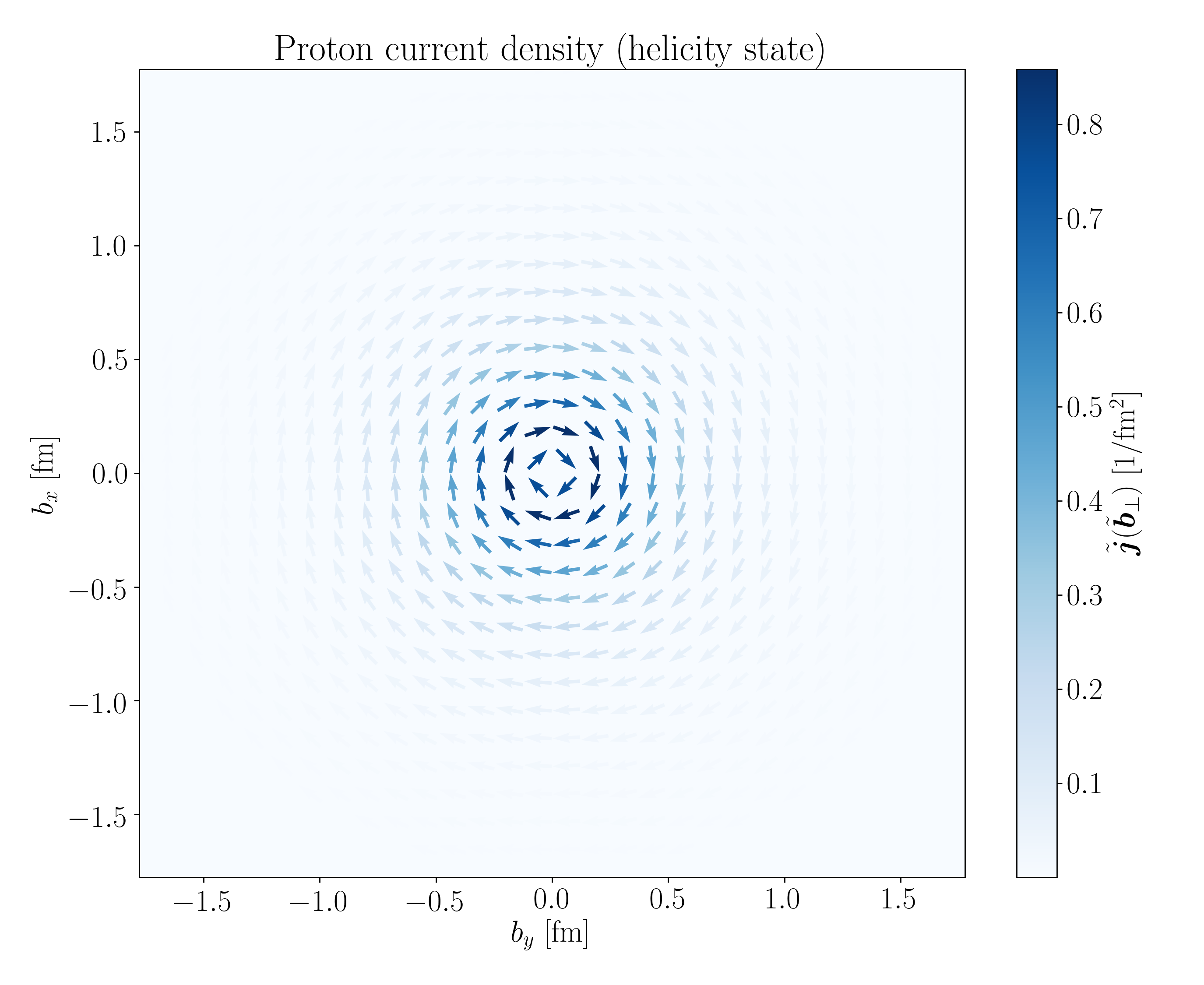}
  \includegraphics[width=0.49\textwidth]{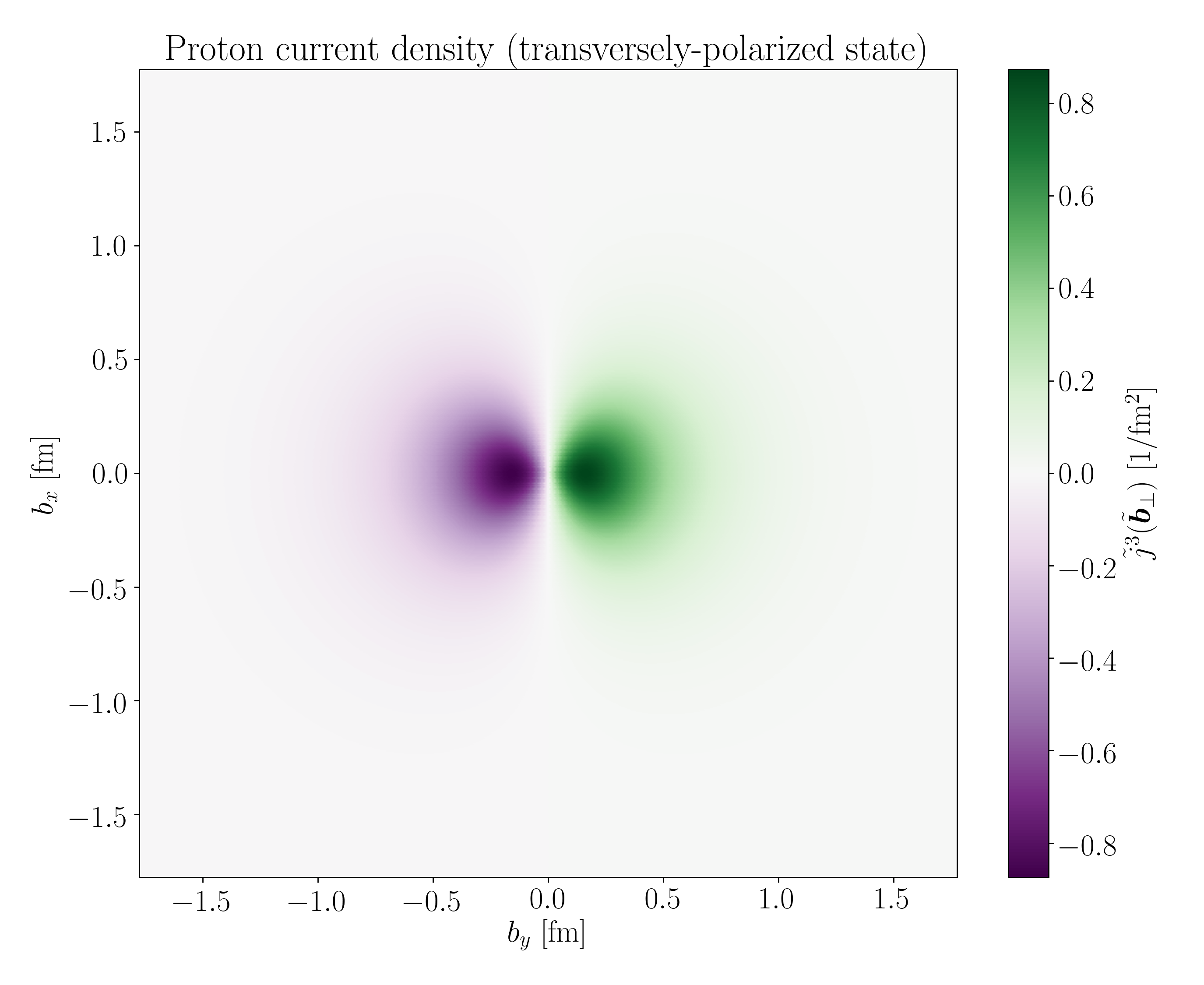}
  \includegraphics[width=0.49\textwidth]{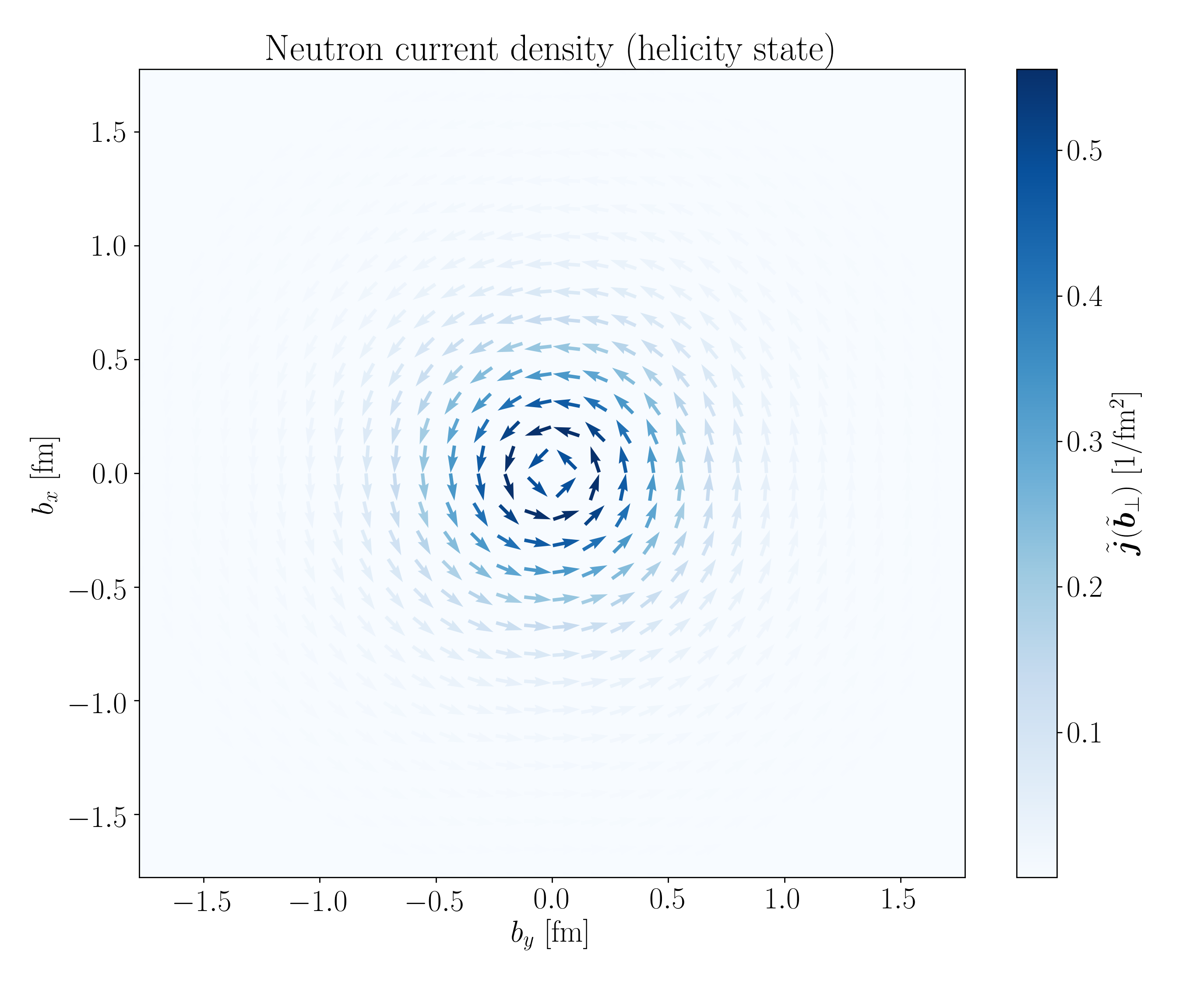}
  \includegraphics[width=0.49\textwidth]{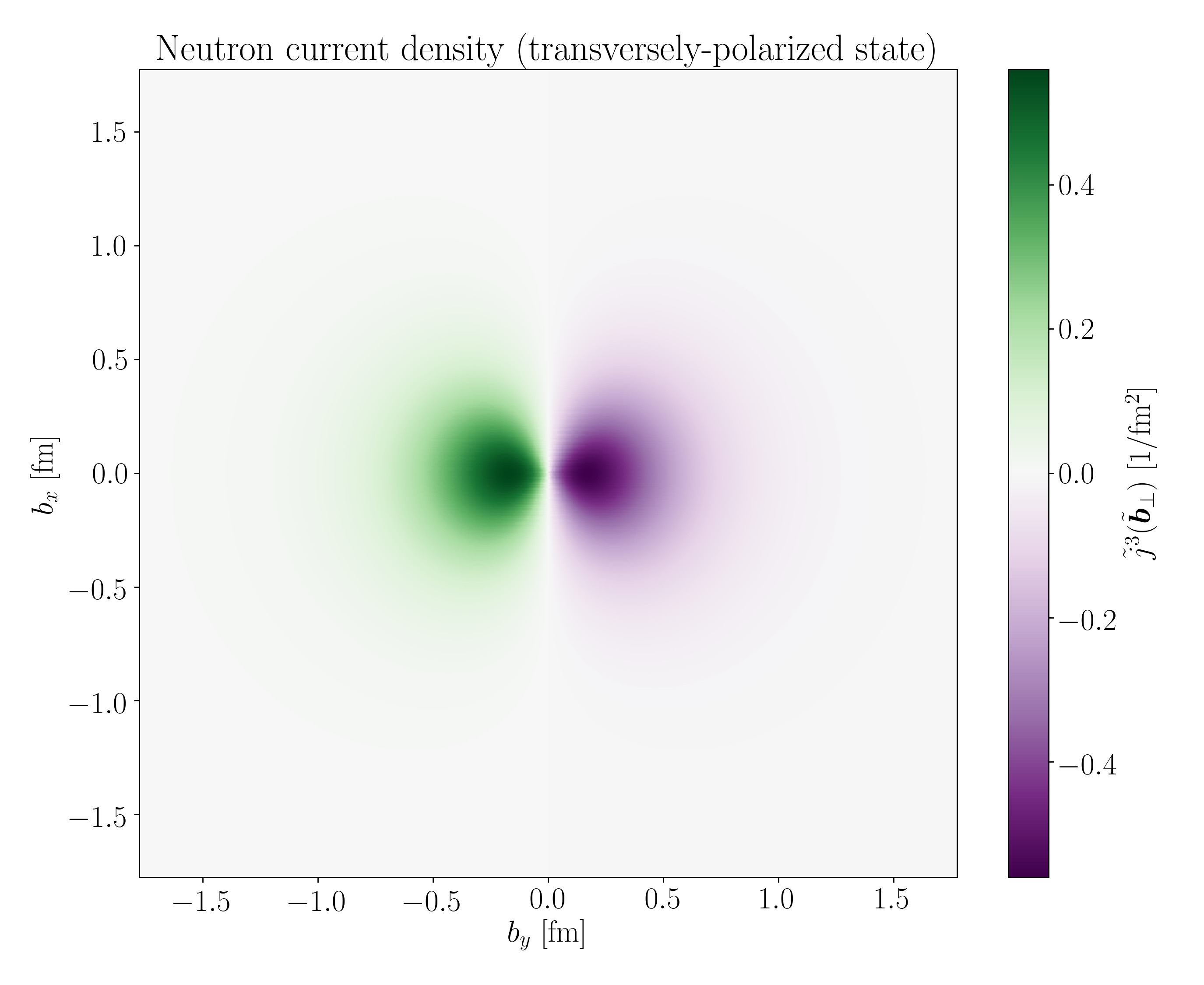}
  \caption{
    Rest frame electric current distributions in the proton (top row) and neutron (bottom row).
    The left column depicts positive-helicity states
    (spin-up along the $z$ axis),
    while the right column depicts transversely-polarized states
    with spin up along the $x$ axis.
    These figures use a right-handed coordinate system with a vertical
    $x$ axis and horizontal $y$ axis, so the $z$ axis points into the page.
  }
  \label{fig:current}
\end{figure}

We next present in Fig.~\ref{fig:current} numerical results for the
internal electric current distributions of the proton and neutron.
It should be noted that we use a right-handed coordinate system in these plots
in which the $x$ axis is oriented vertically and the $y$ axis is oriented
horizontally, so that the $z$ axis points into the page.
This choice is made so that the distributions reflect what an observer
located downstream from the target (i.e., at a lesser $z$ coordinate)
actually sees.
The clockwise revolution of the electric current for a proton with
spin up along the $z$ axis (top left panel) thus entails a positive
magnetic moment, while the counterclockwise revolution of a neutron
in the same state (bottom left panel) entails a negative magnetic moment.
The helicity-state results in the left column reproduce the findings
of Chen and Lorc\'{e}~\cite{Chen:2022smg}.

For a transversely polarized nucleon,
the only non-zero component of its internal current density---as
defined in Eq.~(\ref{eqn:current:internal})---is
$\tilde{j}^3(\tilde{\bm{b}}_\perp)$.
This is depicted in the right column of Fig.~\ref{fig:current}.
That the $z$ axis points into the page is worth emphasizing when interpreting
these plots.
For a proton that is spin-up along the $x$ axis,
the current is moving towards the observer at $b_y < 0$
and away from the observer at $b_y > 0$.
Since the observer looks in the $+z$ direction,
a current towards the observer is negative.
Using the right-hand rule,
the transversely-polarized proton (top-right panel)
has an electric current that co-revolves with its spin,
while the transversely-polarized neutron (bottom-right panel)
has an electric current that counter-rotates against its spin.
These are consistent with the picture suggested by the helicity-state nucleons.

It is worth remarking that the
$\tilde{j}^3(\tilde{\bm{b}}_\perp)$
results for transversely-polarized states are new,
and could be obtained and meaningfully interpreted by virtue of using
tilted coordinates instead of standard light front coordinates.


\subsection{Quark densities and currents}

Information about the densities and currents of up quarks and down quarks
in the proton can be found by invoking charge symmetry.
The $u$ and $d$ densities in the proton are assumed equal to
the $d$ and $u$ densities in the neutron.
Thus, for any density $\rho(\tilde{\bm{b}}_\perp)$,
\begin{subequations}
  \begin{align}
    \rho_p(\tilde{\bm{b}}_\perp)
    &=
    e_u
    \rho_u(\tilde{\bm{b}}_\perp)
    +
    e_d
    \rho_d(\tilde{\bm{b}}_\perp)
    =
    \frac{2}{3}
    \rho_u(\tilde{\bm{b}}_\perp)
    -
    \frac{1}{3}
    \rho_d(\tilde{\bm{b}}_\perp)
    \\
    \rho_n(\tilde{\bm{b}}_\perp)
    &=
    e_u
    \rho_d(\tilde{\bm{b}}_\perp)
    +
    e_d
    \rho_u(\tilde{\bm{b}}_\perp)
    =
    \frac{2}{3}
    \rho_d(\tilde{\bm{b}}_\perp)
    -
    \frac{1}{3}
    \rho_u(\tilde{\bm{b}}_\perp)
    \,,
  \end{align}
\end{subequations}
which invert to:
\begin{subequations}
  \begin{align}
    \rho_u(\tilde{\bm{b}}_\perp)
    &=
    2
    \rho_p(\tilde{\bm{b}}_\perp)
    +
    \rho_n(\tilde{\bm{b}}_\perp)
    \\
    \rho_d(\tilde{\bm{b}}_\perp)
    &=
    2
    \rho_n(\tilde{\bm{b}}_\perp)
    +
    \rho_p(\tilde{\bm{b}}_\perp)
    \,.
  \end{align}
\end{subequations}
The up and down quark densities in a helicity-state proton
were found previously by Miller~\cite{Miller:2007uy}.
We are now equipped to additionally provide quark densities in
transversely-polarized protons,
as well as the quark current densities.

\begin{figure}
  \includegraphics[width=0.49\textwidth]{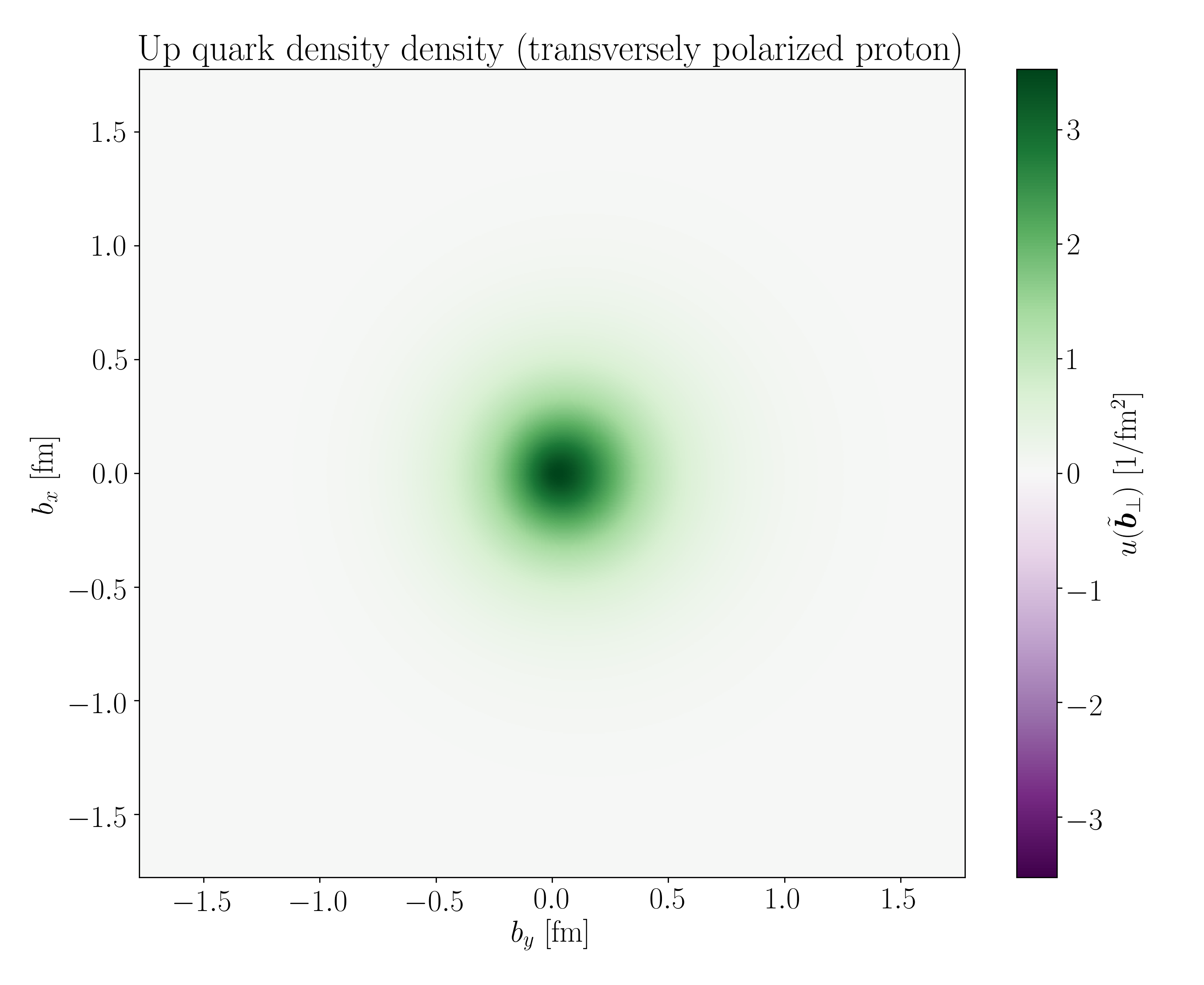}
  \includegraphics[width=0.49\textwidth]{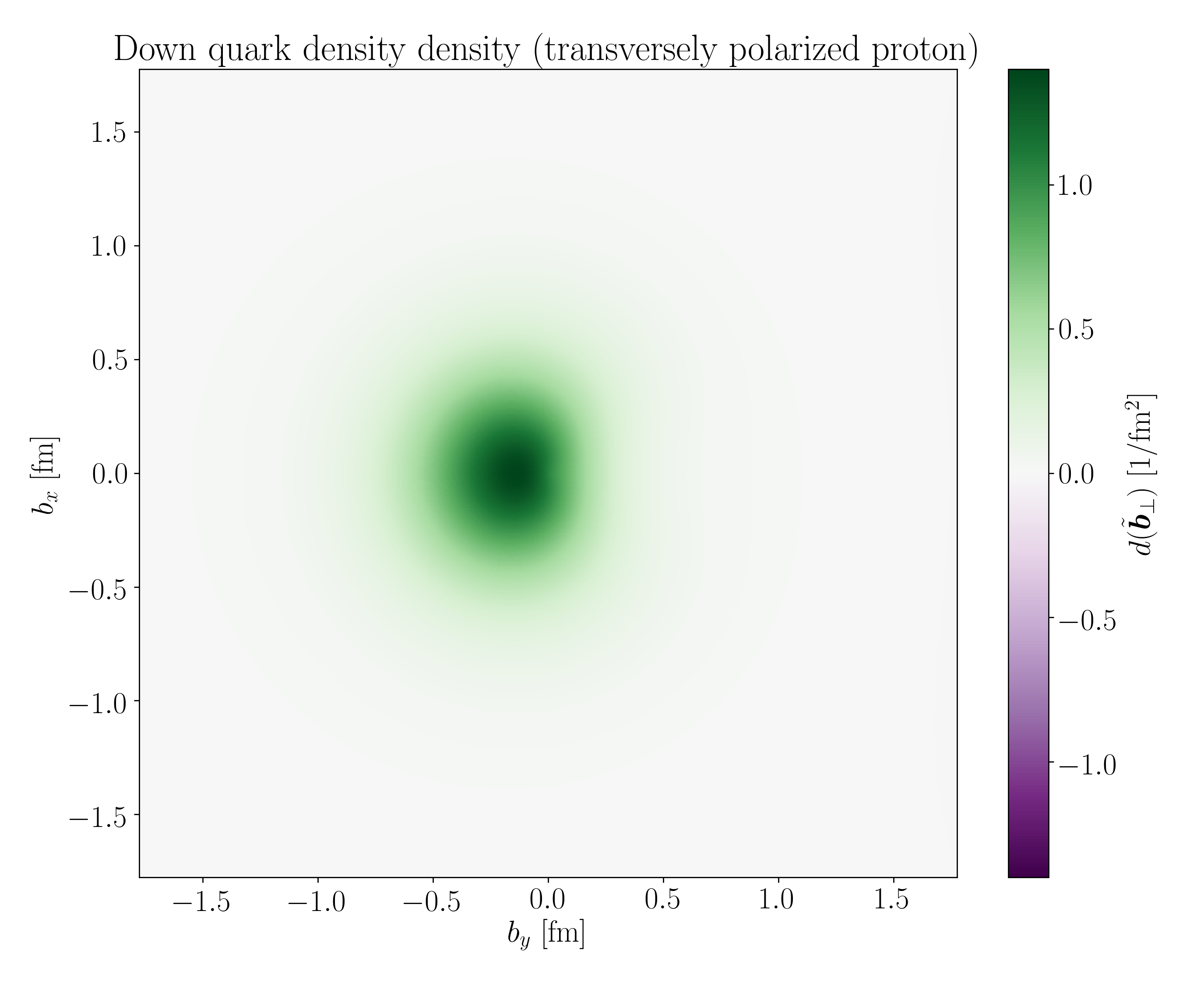}
  \caption{
    Rest frame quark densities in a transversely-polarized proton.
    Rest frame electric densities of the proton (top row) and neutron (bottom row).
    The proton has spin-up along the $x$ axis.
    The left panel is the up quark density and the right panel is the down quark density.
  }
  \label{fig:quark:density}
\end{figure}

Up and down quark densities in a transversely-polarized proton at rest
are presented in Fig.~\ref{fig:quark:density}.
These densities are not weighted by charge, so they are positive-definite
number densities, normalized to $n_u = 2$ and $n_d = 1$ respectively.
Both densities are off-center, but the down quark density is much further
displaced from the center.
These displacements are in opposite directions,
and since the up and down quarks have opposite charge,
they make constructive contributions to the synchronization-induced electric dipole moment.

\begin{figure}
  \includegraphics[width=0.49\textwidth]{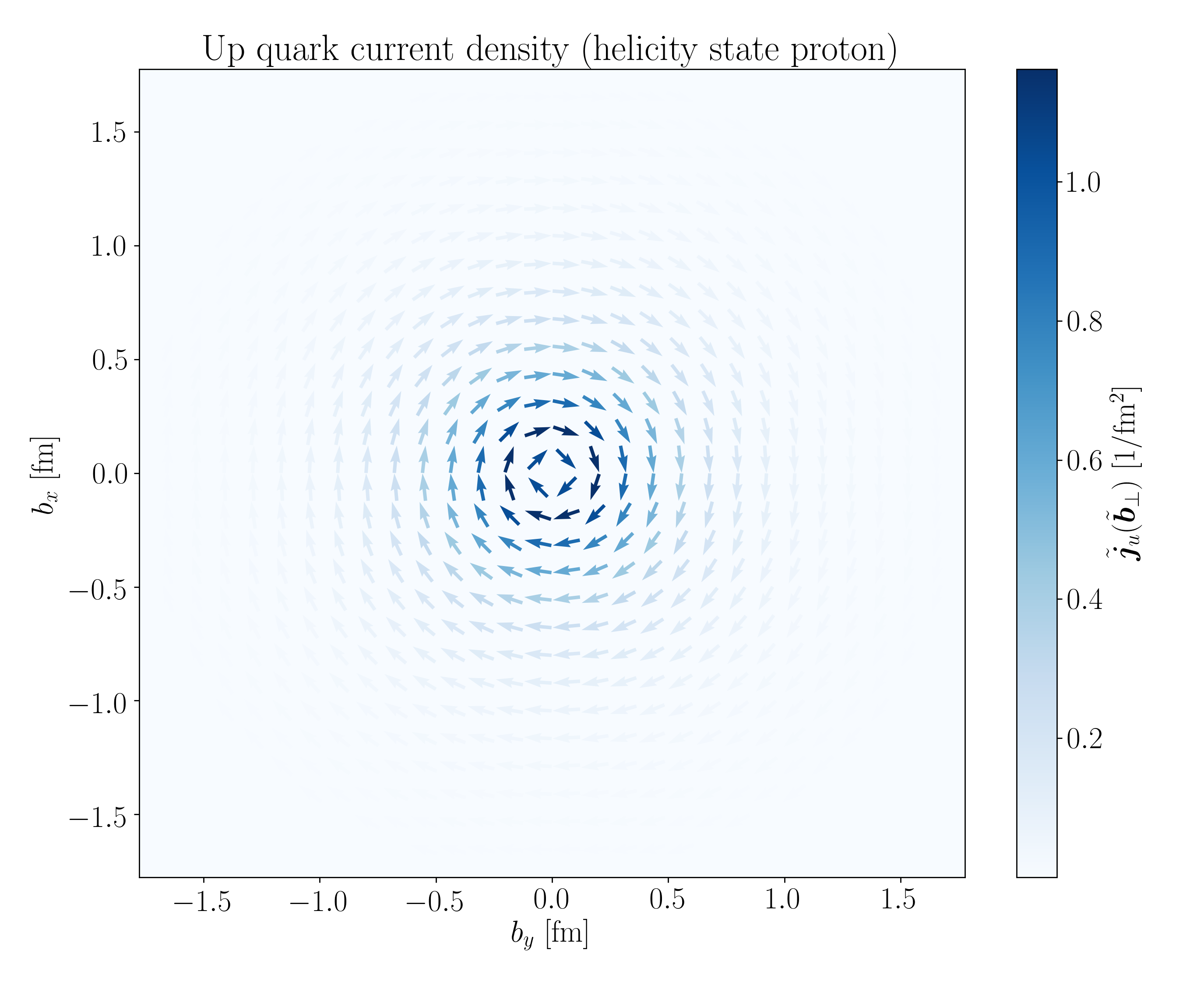}
  \includegraphics[width=0.49\textwidth]{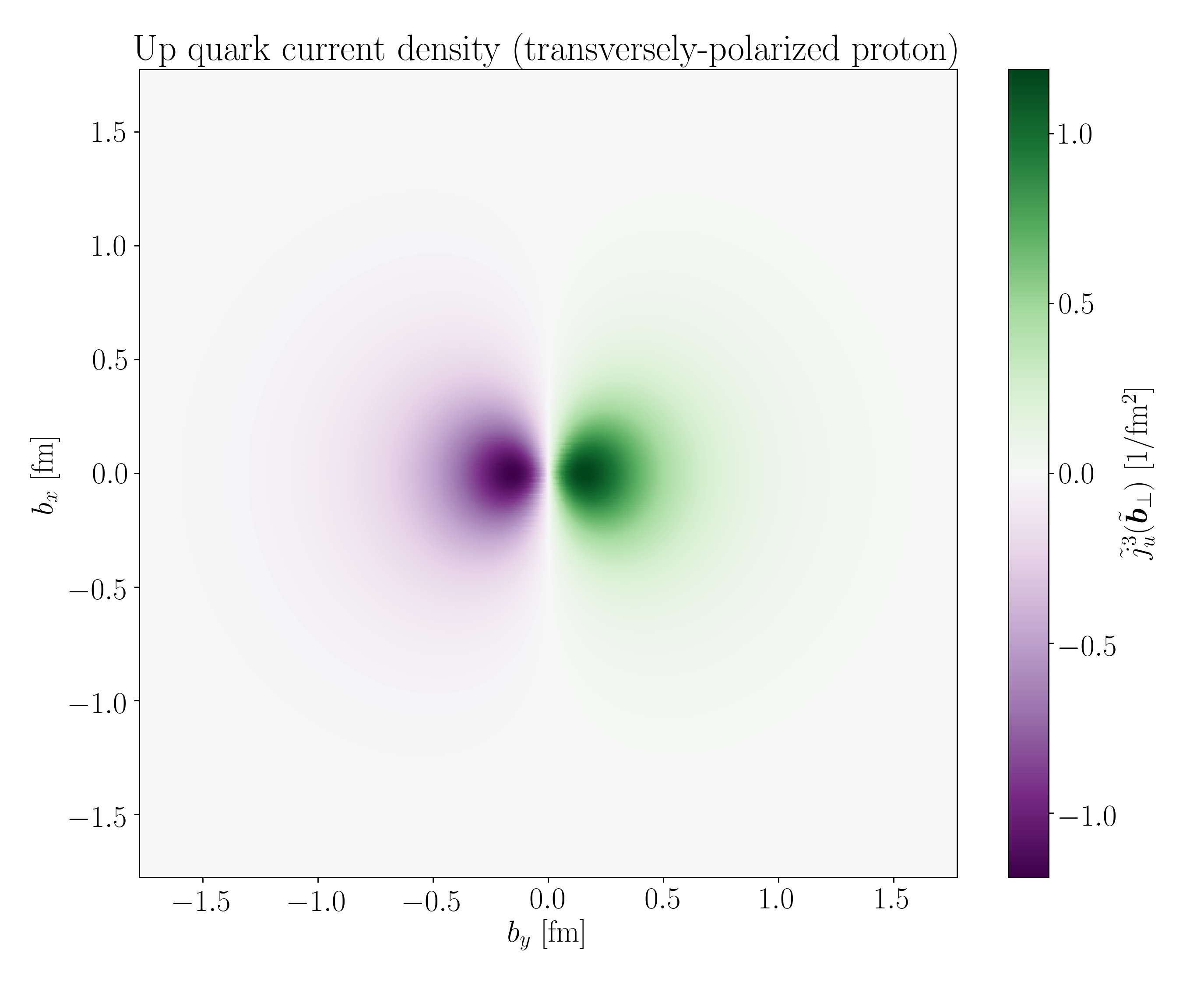}
  \includegraphics[width=0.49\textwidth]{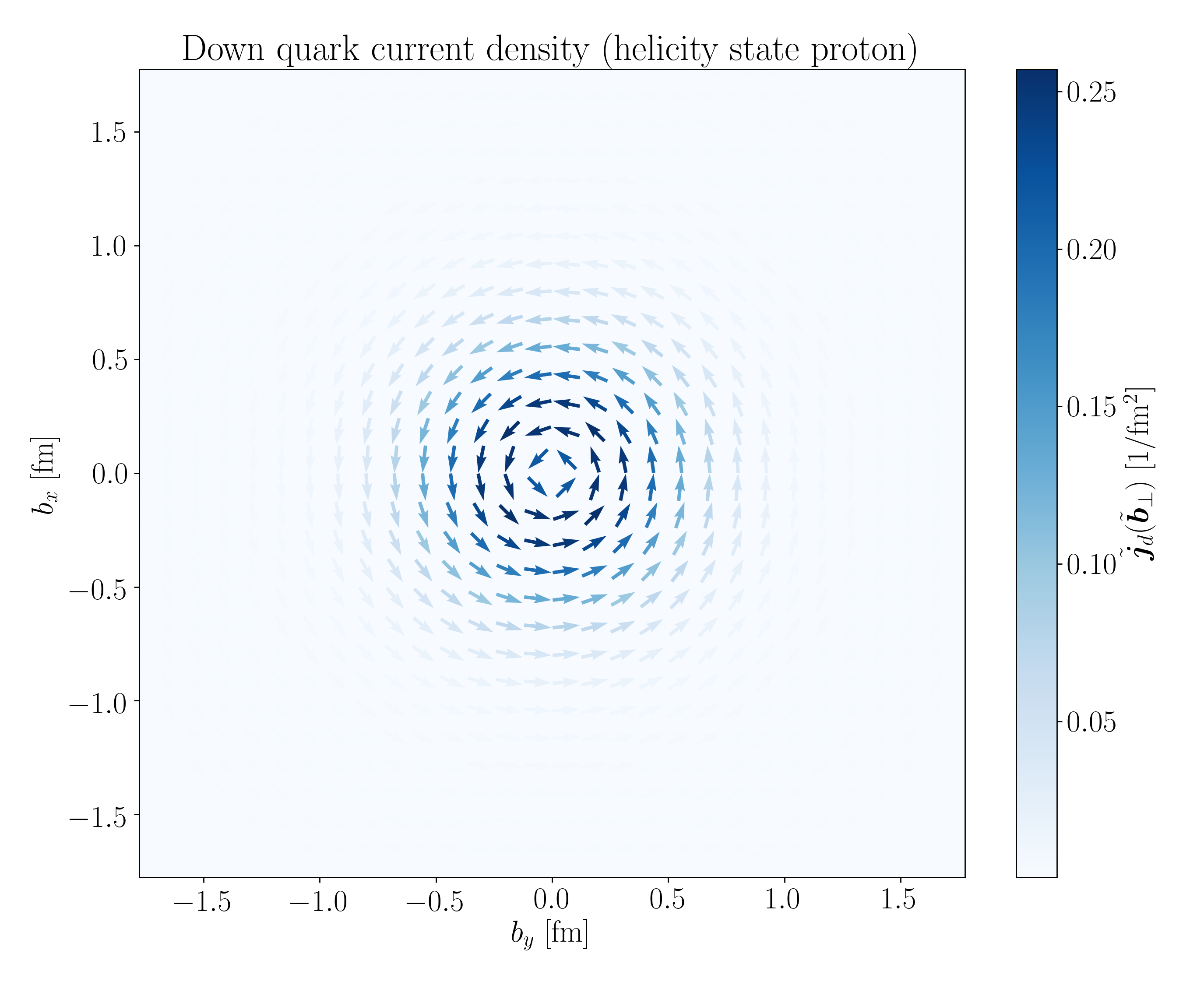}
  \includegraphics[width=0.49\textwidth]{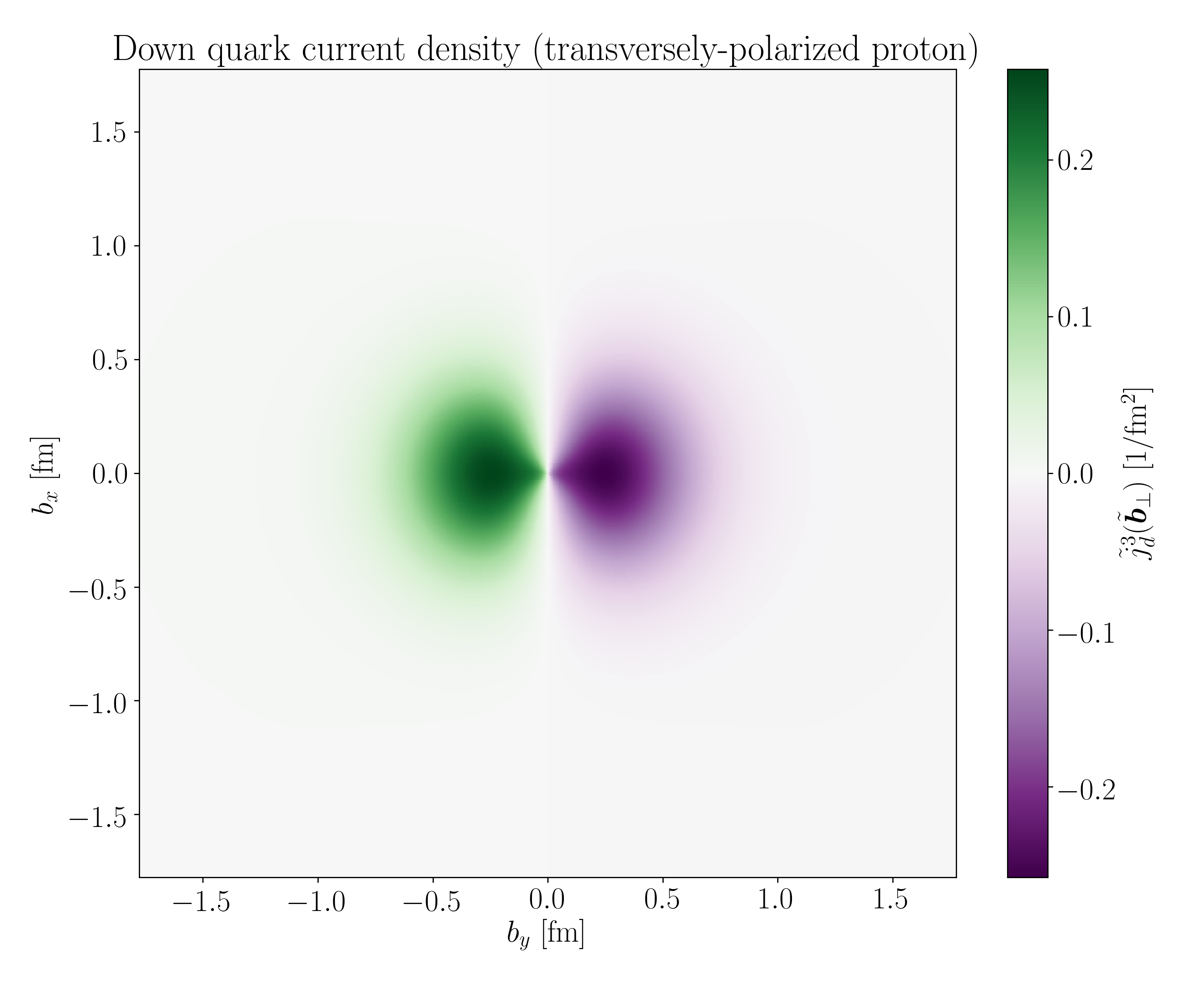}
  \caption{
    Rest frame quark current distributions in the proton.
    Up distributions (top row) and down distributions (bottom row) are shown.
    The left column depicts positive-helicity states
    (spin-up along the $z$ axis),
    while the right column depicts transversely-polarized states
    with spin up along the $x$ axis.
    These figures use a right-handed coordinate system with a vertical
    $x$ axis and horizontal $y$ axis, so the $z$ axis points into the page.
  }
  \label{fig:quark:current}
\end{figure}

Up and down quark current distributions in a proton
are presented in Fig.~\ref{fig:quark:current}.
These are not weighted by charge, so the direction of the current is the genuine
direction of motion of the respective quark flavor.
For both the longitudinally and transversely-polarized state,
the up quark current is in the direction of proton spin,
while the down quark current is in the opposite direction.
This seems to conflict with model calculations~\cite{Thomas:2008ga,Freese:2020mcx}
and lattice QCD results~\cite{Alexandrou:2017oeh} finding
that up quarks carry negative orbital angular momentum (OAM)
and down quarks carry positive OAM,
at least at renormalization scales of a few GeV$^2$ or higher.
However, the OAM contains sea quark contributions---hence
its renormalization scale dependence---while
the quark currents describe only valence quark contributions.
This conflict is resolved if the majority of OAM is carried by sea quarks.

We remark, in comparing Fig.~\ref{fig:quark:density} to Fig.~\ref{fig:quark:current},
that each quark density appears to be enhanced where the quarks are moving away
from the observer
(on the right side of the transversely-polarized plots of up quark distributions,
or the left side for down quark distributions),
and conversely suppressed where the quarks are moving towards the observer.
This is a natural consequence of using light front time synchronization,
and will be explored further in Sec.~\ref{sec:discuss}.

\begin{table}[t]
  \setlength{\tabcolsep}{0.5em}
  \renewcommand{\arraystretch}{1.3}
  \begin{tabular}{@{} cccccc @{}}
    \toprule
    $ n_u $ &
    $ n_d $ &
    $ \langle (\tilde{\bm{b}}_\perp^2)_u \rangle $ &
    $ \langle (\tilde{\bm{b}}_\perp^2)_d \rangle $ &
    $ \kappa_u $ &
    $ \kappa_d $ \\
    \hline
    2 &
    1 &
    $\big(0.65~\text{fm}\big)^2$ &
    $\big(0.68~\text{fm}\big)^2$ &
    $1.68$ &
    $-2.03$ \\
    \bottomrule
  \end{tabular}
  \caption{
    Static quantities associated with up and down quarks in the proton.
  }
  \label{tab:ud}
\end{table}

In Table~\ref{tab:ud}, we present individual quark flavors' contributions
to the static quantities associated with the proton's charge and current distributions.
The quark number, quark radius, and quark contribution to the anomalous magnetic moment
(without charge weight) are respectively:
\begin{subequations}
  \begin{align}
    n_q
    &=
    F_{1q}(0)
    \\
    \langle (\tilde{\bm{b}}_\perp^2)_q \rangle
    &=
    \frac{4}{F_{1q}(0)} \frac{\d F_{1q}(t)}{\d t}\bigg|_{t=0}
    \\
    \kappa_q
    &=
    F_{2q}(0)
    \,.
  \end{align}
\end{subequations}


\subsection{Discussion and further interpretation}
\label{sec:discuss}

We now further discuss the interpretation of the internal densities we have
obtained, with a special focus on the angular modulations in the charge
densities of transversely-polarized nucleons
(right column of Fig.~\ref{fig:charge}).
Since these densities are frame-invariant and therefore describe
a proton even at rest, they cannot be attributed to kinematic effects
from boosting to infinite momentum.

Modulations such as these can appear for spinning targets at rest by virtue of
using light front synchronization.
What a downstream observer sees at any fixed light front time $\tilde{\tau}$
is everything that a single light front encounters while traveling
in the $-z$ direction.
For an extended rotating body, the light front takes a non-zero amount of
Einstein time $t$ to cross the extent of the body,
and particles within it will advance in their motion while the light front
is crossing it.
Particles are more likely to encounter the light front while they're
moving against it (in the $+z$ direction),
so the particle density will be skewed in the $\hat{z}\times\hat{\bm{s}}$
direction.
This can explain the direction of the modulations in of Figs.~\ref{fig:charge} and \ref{fig:quark:density},
as well as the sign of the synchronization-induced electric dipole moment
in Eq.~(\ref{eqn:dipole}).

This same phenomenon can also be rephrased in the language of light front time $\tilde{\tau}$.
Suppose that a particle revolves around a transversely-polarized body,
taking the same amount of time from its own perspective to reach the lowest-$z$
point of its orbit and the highest-$z$ point of its orbit.
Since its time is dilated when moving in the $+z$ direction and quickened
when moving in the $-z$ direction from the perspective of a stationary observer
(see discussion in Sec.~\ref{sec:lorentz}),
it will spend less time moving towards the observer and more time moving away
from the observer from the observer's perspective.
This makes it more likely that the observer will see the particle when it's
in the $\hat{z}\times\hat{\bm{s}}$ direction,
again skewing the particle density in that direction.

This can be explicitly illustrated in a simple classical toy model.
Suppose we have a charged body revolving around a central point
with angular velocity $\omega$.
The revolution is around the $x$-axis,
and the trajectory in terms of Einstein time $t$ is:
\begin{align}
  \bm{r}(t)
  &=
  R \Big(
  \hat{y}\cos(\omega t)
  +
  \hat{z}\sin(\omega t)
  \Big)
  \,.
\end{align}
At any Einstein time, the $z$ coordinate is:
\begin{align}
  z(t)
  &=
  \hat{z}\cdot \bm{r}(t)
  =
  R \sin(\omega t)
  \,.
\end{align}
The light front time $\tilde{\tau}$ and Einstein time are related by:
\begin{align}
  \tilde{\tau}
  =
  t
  +
  \frac{z(t)}{c}
  \equiv
  t
  +
  \frac{\tilde{z}(\tilde{\tau})}{c}
  \,,
\end{align}
so that the $z$ coordinate as a function of the light front time $\tilde{\tau}$
is given by the implicit formula:
\begin{align}
  \tilde{z}(\tilde{\tau})
  =
  R \sin\big(\omega [\tilde{\tau} - \tilde{z}(\tilde{\tau})]\big)
  \,.
\end{align}
The function $\tilde{z}(\tilde{\tau})$ can be solved for numerically.
Additionally, from this one can find the $y$ coordinate as a function of light front
time from this:
\begin{align}
  \tilde{y}(\tilde{\tau})
  &=
  y(t)
  =
  R \cos\big(\omega [\tilde{\tau} - \tilde{z}(\tilde{\tau})]\big)
  \,.
\end{align}

\begin{figure}
  \includegraphics[width=0.49\textwidth]{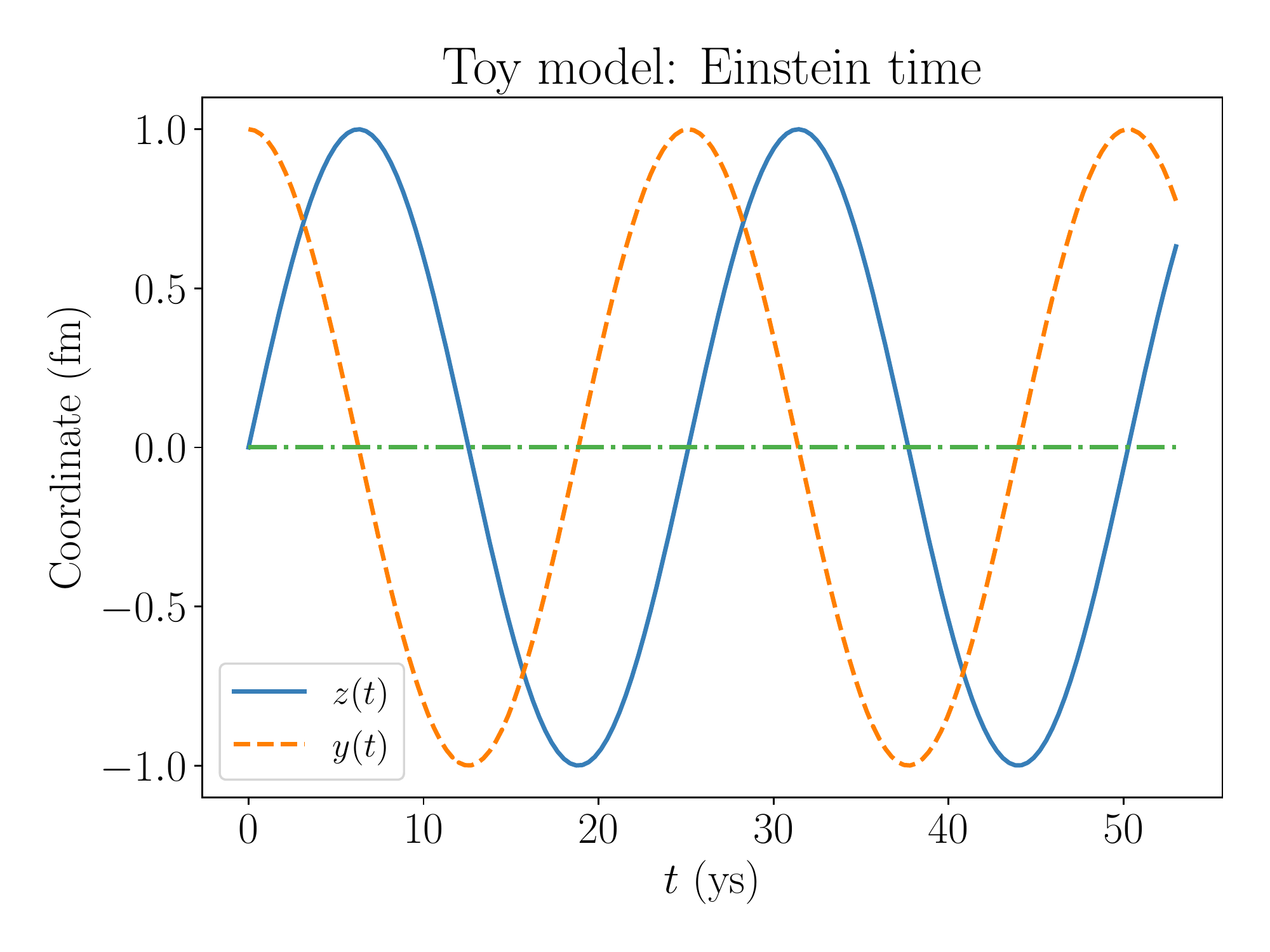}
  \includegraphics[width=0.49\textwidth]{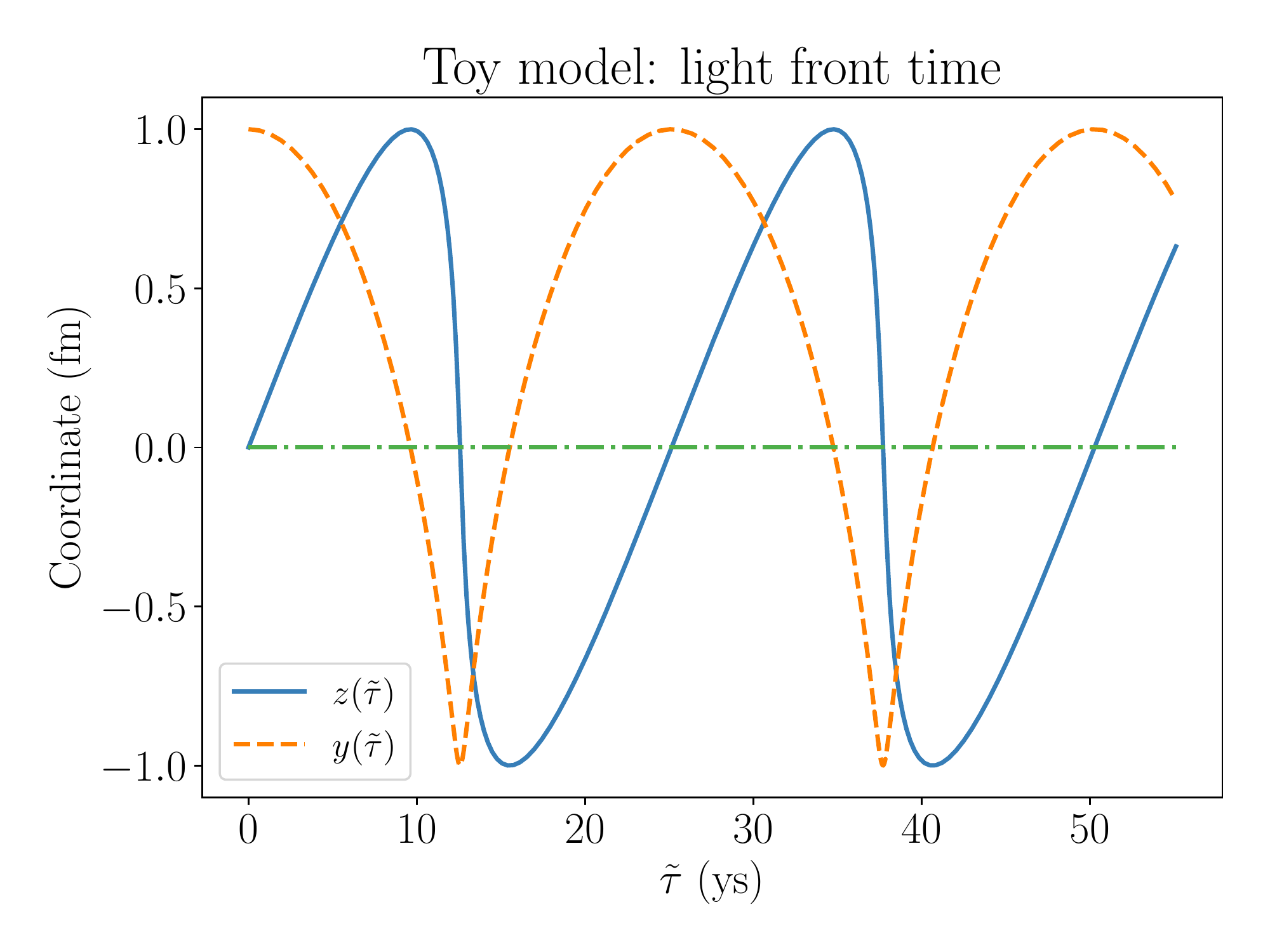}
  \caption{
    $y$ and $z$ coordinates of the toy revolver model as functions of
    Einstein time (left panel) and light front time (right panel),
    for $R = 1$~fm and $\omega = 250$~ZHh.
  }
  \label{fig:toy}
\end{figure}

Numerical results for the revolver coordinates as a function of light front time $\tilde{\tau}$
are presented in the right panel of Fig.~\ref{fig:toy},
with the Einstein time trajectory given in the left panel for contrast.
In terms of Einstein time $t$, the coordinates are simple sinusoids, as expected.
In terms of light front time, however, the trajectory appears highly distorted.
The revolver spends most of its time moving away from the observer with a low velocity,
and then quickly swings back towards the observer at a high velocity in a short span of light front time.
Because of this, the revolver is at a positive $y$ coordinate for longer than
it is at a negative $y$ coordinate,
and the light-front-time-averaged $y$ coordinate is non-zero.
In fact, this average can be calculated analytically:
\begin{align}
  \langle \tilde{y}(\tilde{\tau}) \rangle_{\text{LF}}
  & \equiv
  \lim_{T\rightarrow\infty}
  \frac{1}{2 \tilde{\tau}(T)}
  \int_{-\tilde{\tau}(T)}^{+\tilde{\tau}(T)}
  \d \tilde{\tau}' \,
  \tilde{y}(\tilde{\tau}')
  =
  \lim_{T\rightarrow\infty}
  \frac{1}{2 T + 2\frac{R}{c}\sin(\omega T)}
  \int_{-T}^{+T}
  \d t \,
  \frac{\d \tilde{\tau}(t)}{\d t}
  y(t)
  =
  \frac{R^2\omega}{2c}
  \,.
\end{align}
We can use this model to give an extremely rough estimate of how quickly
the valence up and down quarks revolve around the proton.
These estimates will be rough because quarks do not orbit
the proton in well-defined circular trajectories,
but they should give an idea of the order of magnitude of the angular
frequency of quark revolutions.
If $n_q \langle \tilde{y}(\tilde{\tau}) \rangle_{\text{LF}}$ is interpreted
as the synchronization-induced dipole moment of the quark distribution,
then using this with Eq.~(\ref{eqn:dipole}) gives:
\begin{align}
  \omega_q
  =
  \frac{c\kappa_q}{m_p n_q R_q^2}
  \,.
\end{align}
For the sake of these estimates, the mean-squared radii given in
Table~\ref{tab:ud} can be used in place of $R_q^2$.
Doing this gives the following rough estimates:
\begin{subequations}
  \begin{align}
    \omega_u
    & \approx
    \phantom{-}
    0.417~c/\text{fm}
    =
    \phantom{-}
    125~\text{ZHz}
    \\
    \omega_d
    & \approx
    -0.922~c/\text{fm}
    =
    -276~\text{ZHz}
    \,,
  \end{align}
\end{subequations}
where the minus sign for the down quark's angular frequency indicates
revolution in the opposite direction.
These exact numbers depended on a simplistic toy model and should be taken
only to give a ballpark estimate of the angular velocity of
up and down quarks' orbits around the proton.


\section{Conclusions and outlook}
\label{sec:end}

In this work, we used tilted light front coordinates---light front time
$x^+$ and ordinary spatial coordinates $(x,y,z)$---to demonstrate that
the formalism for light front densities follows from use of a peculiar
time synchronization convention,
rather than from boosting to infinite momentum.
Internal densities of electric charge and currents,
as well as up and down quark densities and currents,
were obtained for a proton.
These densities are boost-invariant,
and thus describe a proton in any state of motion,
including at rest.
The densities and transverse currents we obtained reproduced prior results,
while the longitudinal current densities were newly-obtained
by virtue of using tilted coordinates.

This formalism is equally applicable to the energy-momentum tensor (EMT),
allowing all sixteen of its components to be given a clear physical meaning.
This will become the subject of a sequel paper.
Care will need to be taken about upper and lower indices;
since the energy and momentum are identified as covariant components of
$\tilde{p}_\mu$ (as discussed in Sec.~\ref{sec:properties}),
and since the EMT is conserved with respect to its
first index, the tilted energy density for instance will be given by:
\begin{align}
  \tilde{T}^0_{\phantom{\mu}0}
  =
  \tilde{T}^{00} - \tilde{T}^{03}
  =
  T^{00} + T^{30}
  \,,
\end{align}
which involves off-diagonal components of the EMT.
For spin-half targets in particular,
the energy density may differ
depending on whether the EMT contains the antisymmetric piece
discussed in Refs.~\cite{Leader:2013jra,Lorce:2018egm}.
Tilted light front coordinates will be helpful not only
in defining relativistically exact densities for all
components of the EMT---including an energy density that
integrates to the usual instant form energy $E$---but
may also provide additional insight into the question
of whether the physical EMT contains an antisymmetric piece.


\begin{acknowledgments}
  The authors would like to thank
  Ian Clo\"{e}t, Wim Cosyn, and Yang Li
  for helpful discussions that contributed to this work.
  This work was supported by the U.S.\ Department of Energy
  Office of Science, Office of Nuclear Physics under Award Number
  DE-FG02-97ER-41014.
\end{acknowledgments}


\appendix

\section{Explicit light front helicity spinors}
\label{sec:spinors}

To evaluate internal proton densities,
we will need to evaluate explicit spinor matrix elements.
To this end, it is helpful to have explicit expressions for
light front helicity spinors.
Explicit expressions for these spinors were found already by
Kogut and Soper~\cite{Kogut:1969xa}---the Kogut-Soper spinors are in fact eigenstates
of the light front helicity operator $\hat{\lambda}$
defined in Eq.~(\ref{eqn:hel:op}).

In terms of tilted momentum components, the Kogut-Soper spinors can be written:
\begin{align}
  &
  u\left(\tilde{p},+\frac{1}{2}\right)
  =
  \frac{1}{\sqrt{\tilde{p}_z}}
  \left[
    \begin{array}{c}
      \tilde{p}_z \\
      \tilde{p}_x + \i\tilde{p}_y \\
      m \\
      0
    \end{array}
    \right]
  \,,
  &
  u\left(\tilde{p},-\frac{1}{2}\right)
  =
  \frac{1}{\sqrt{\tilde{p}_z}}
  \left[
    \begin{array}{c}
      0 \\
      m \\
      -\tilde{p}_x + \i\tilde{p}_y \\
      \tilde{p}_z
    \end{array}
    \right]
  \,.
\end{align}
In this basis, the tilted gamma matrices can be written in block matrix form as:
\begin{align}
  &
  \tilde{\gamma}^0
  =
  \left[
    \begin{array}{cc}
      0 & \mathbbm{1}-\sigma_3 \\
      \mathbbm{1}+\sigma_3 & 0
    \end{array}
    \right]
  \,,
  &
  \tilde{\bm{\gamma}}^i
  =
  \left[
    \begin{array}{cc}
      0 & -\sigma_i \\
      \sigma_i & 0
    \end{array}
    \right]
  \,,
\end{align}
where $\mathbbm{1}$ is a $2\times 2$ identity matrix
and $\sigma_i$ are the usual $2\times 2$ Pauli matrices.

The barred spinors are defined as usual by:
\begin{subequations}
  \begin{align}
    \bar{u}(\tilde{\bm{p}},\lambda)
    &=
    u^\dagger(\tilde{\bm{p}},\lambda) \gamma^0
    \\
    \bar{v}(\tilde{\bm{p}},\lambda)
    &=
    v^\dagger(\tilde{\bm{p}},\lambda) \gamma^0
  \end{align}
\end{subequations}
where the untilted $\gamma^0$ matrix is given by:
\begin{align}
  \gamma^0
  =
  \left[
    \begin{array}{cc}
      0 & \mathbbm{1} \\
      \mathbbm{1} & 0
    \end{array}
    \right]
  \,.
\end{align}
The reason that $\gamma^0$ rather than $\tilde{\gamma}^0$ is used
is that $\gamma^0$ plays the role here of swapping
left-handed and right-handed components of the spinor,
rather than of isolating a component of a four-vector.
Indeed, since
$\bar{u}(\tilde{\bm{p}}',\lambda') u(\tilde{\bm{p}},\lambda)
=
u^\dagger(\tilde{\bm{p}}',\lambda') \gamma^0 u(\tilde{\bm{p}},\lambda)$
is a scalar, it should be invariant under the transformation from
Minkowski to tilted coordinates.

It is helpful to have explicit matrix elements of spinors,
in particular when momentum in the $z$ direction remains unchanged.
Let $\tilde{\bm{p}}  = \tilde{\bm{P}} - \frac{1}{2}\tilde{\bm{\varDelta}}$
and $\tilde{\bm{p}}' = \tilde{\bm{P}} + \frac{1}{2}\tilde{\bm{\varDelta}}$,
and $\tilde{\bm{\varDelta}}_z=0$.
For equal-helicity spinors, we find the following matrix elements:
\begin{subequations}
  \begin{align}
    \bar{u}(\tilde{\bm{p}}',\lambda)
    u(\tilde{\bm{p}},\lambda)
    &=
    2m
    \\
    \bar{u}(\tilde{\bm{p}}',\lambda)
    \tilde{\gamma}^0
    u(\tilde{\bm{p}},\lambda)
    &=
    2 \tilde{P}_z
    \\
    \bar{u}(\tilde{\bm{p}}',\lambda)
    \tilde{\gamma}_\perp^i
    u(\tilde{\bm{p}},\lambda)
    &=
    2 \tilde{P}^i_\perp
    -
    2 \lambda
    \i \epsilon_{ij3} \tilde{\varDelta}^j
    \\
    \bar{u}(\tilde{\bm{p}}',\lambda)
    \tilde{\gamma}^3
    u(\tilde{\bm{p}},\lambda)
    &=
    2 \tilde{P}_z
    \left(1 - \frac{\tilde{E}+\tilde{E}'}{2\tilde{P}_z}\right)
    +
    \frac{
      2\lambda
      \i
      (\tilde{\bm{P}}_\perp \times \tilde{\bm{\varDelta}}_\perp)\cdot\hat{z}
    }{\tilde{P}_z}
    \,,
  \end{align}
\end{subequations}
where $\tilde{E}$ and $\tilde{E}'$ are the mass-shell energies
of $\tilde{\bm{p}}$ and $\tilde{\bm{p}}'$, respectively.
The gamma matrix elements in particular can be summarized
in the following manifestly-covariant form:
\begin{align}
  \bar{u}(\tilde{\bm{p}}',\lambda)
  \tilde{\gamma}^\mu
  u(\tilde{\bm{p}},\lambda)
  &=
  2 \tilde{P}^\mu
  -
  \frac{
    2 \lambda
    \i
    \epsilon^{\mu\nu\rho\sigma}
    \tilde{n}_\nu
    \tilde{P}_\rho \tilde{\varDelta}_\sigma
  }{(\tilde{P}\cdot\tilde{n})}
  \,,
\end{align}
where $\tilde{n}_\mu = (1;0,0,0)$ projects out the
``time'' component of contravariant four-vectors
(e.g., $\tilde{n}_\mu \tilde{x}^\mu = \tilde{x}^0 = \tilde{\tau}$).

Let us consider the helicity-flip case next.
For $\lambda'\neq\lambda$, let $\Delta\lambda = \lambda'-\lambda$.
We find:
\begin{subequations}
  \begin{align}
    \bar{u}(\tilde{\bm{p}}',\lambda')
    u(\tilde{\bm{p}},\lambda)
    &=
    \big(
    \Delta\lambda \tilde{\varDelta}_x
    - \i \tilde{\varDelta}_y
    \big)
    \\
    \bar{u}(\tilde{\bm{p}}',\lambda')
    \tilde{\gamma}^0
    u(\tilde{\bm{p}},\lambda)
    &=
    0
    \\
    \bar{u}(\tilde{\bm{p}}',\lambda')
    \tilde{\gamma}_\perp^i
    u(\tilde{\bm{p}},\lambda)
    &=
    0
    \\
    \bar{u}(\tilde{\bm{p}}',\lambda')
    \tilde{\gamma}^3
    u(\tilde{\bm{p}},\lambda)
    &=
    -
    \frac{m}{\tilde{P}_z}
    \big(
    \Delta\lambda \tilde{\varDelta}_x
    - \i \tilde{\varDelta}_y
    \big)
    \,.
  \end{align}
\end{subequations}
Writing these in a manifestly-covariant manner is trickier,
but can be done by utilizing the four-vector
$\tilde{\bar{n}}^\mu = (1;0,0,0)$ which projects out the ``time''
component of covariant four-vectors
(e.g., $\tilde{\bar{n}}^\mu \tilde{p}_\mu = \tilde{p}_0 = \tilde{E}$).
Using Eq.~(\ref{eqn:covcon}), we note that
$\tilde{n}^\mu = (0;0,0,-1)$ and $\tilde{\bar{n}}_\mu = (1;,0,0,-1)$.
With this in mind, we can write:
\begin{subequations}
  \begin{align}
    \bar{u}(\tilde{\bm{p}}',\lambda')
    u(\tilde{\bm{p}},\lambda)
    &=
    \i
    \epsilon^{\alpha\beta\gamma\delta}
    \tilde{n}_\alpha \tilde{\bar{n}}_\beta
    \tilde{\varDelta}_\gamma (\sigma_\delta)_{\lambda'\lambda}
    \\
    \bar{u}(\tilde{\bm{p}}',\lambda')
    \tilde{\gamma}^\mu
    u(\tilde{\bm{p}},\lambda)
    &=
    \frac{m\tilde{n}^\mu}{(\tilde{P}\cdot\tilde{n})}
    \i
    \epsilon^{\alpha\beta\gamma\delta}
    \tilde{n}_\alpha \tilde{\bar{n}}_\beta
    \tilde{\varDelta}_\gamma (\sigma_\delta)_{\lambda'\lambda}
  \end{align}
\end{subequations}
for $\lambda'\neq\lambda$,
where $\sigma_\delta = (\mathbbm{1};\sigma_1,\sigma_2,\sigma_3)$
is a covariant four-vector of Pauli matrices.

The definite-helicity and helicity-flip cases can now be combined:
\begin{subequations}
  \begin{align}
    \bar{u}(\tilde{\bm{p}}',\lambda')
    u(\tilde{\bm{p}},\lambda)
    &=
    2m (\sigma_0)_{\lambda'\lambda}
    +
    \i
    \epsilon^{\alpha\beta\gamma\delta}
    \tilde{n}_\alpha \tilde{\bar{n}}_\beta
    \tilde{\varDelta}_\gamma (\sigma_\delta)_{\lambda'\lambda}
    \\
    \bar{u}(\tilde{\bm{p}}',\lambda')
    \tilde{\gamma}^\mu
    u(\tilde{\bm{p}},\lambda)
    &=
    2 \tilde{P}^\mu
    (\sigma_0)_{\lambda'\lambda}
    -
    \frac{
      \i
      \epsilon^{\mu\nu\rho\sigma}
      \tilde{n}_\nu
      \tilde{P}_\rho \tilde{\varDelta}_\sigma
    }{(\tilde{P}\cdot\tilde{n})}
    (\sigma_3)_{\lambda'\lambda}
    +
    \frac{m\tilde{n}^\mu}{(\tilde{P}\cdot\tilde{n})}
    \i
    \epsilon^{\alpha\beta\gamma\delta}
    \tilde{n}_\alpha \tilde{\bar{n}}_\beta
    \tilde{\varDelta}_\gamma (\sigma_\delta)_{\lambda'\lambda}
    \,.
  \end{align}
  Using Gordon decomposition, we can also write:
  \begin{align}
    \bar{u}(\tilde{\bm{p}}',\lambda')
    \frac{\i \sigma^{\mu\nu} \tilde{\varDelta}_\nu}{2m}
    u(\tilde{\bm{p}},\lambda)
    &=
    -
    \frac{
      \i
      \epsilon^{\mu\nu\rho\sigma}
      \tilde{n}_\nu
      \tilde{P}_\rho \tilde{\varDelta}_\sigma
    }{(\tilde{P}\cdot\tilde{n})}
    (\sigma_3)_{\lambda'\lambda}
    +
    \left(
    \frac{m\tilde{n}^\mu}{(\tilde{P}\cdot\tilde{n})}
    -
    \frac{\tilde{P}^\mu}{m}
    \right)
    \i
    \epsilon^{\alpha\beta\gamma\delta}
    \tilde{n}_\alpha \tilde{\bar{n}}_\beta
    \tilde{\varDelta}_\gamma (\sigma_\delta)_{\lambda'\lambda}
    \,.
  \end{align}
\end{subequations}
For the sake of explicit evaluations, it is helpful to note:
\begin{subequations}
  \label{eqn:helpful}
  \begin{align}
    -
    \frac{
      \i
      \epsilon^{0\nu\rho\sigma}
      \tilde{n}_\nu
      \tilde{P}_\rho \tilde{\varDelta}_\sigma
    }{(\tilde{P}\cdot\tilde{n})}
    &=
    0
    \\
    -
    \frac{
      \i
      \epsilon^{i\nu\rho\sigma}
      \tilde{n}_\nu
      \tilde{P}_\rho \tilde{\varDelta}_\sigma
    }{(\tilde{P}\cdot\tilde{n})}
    &=
    \i (\hat{z} \times \tilde{\bm{\varDelta}}_\perp)_i
    \qquad : i=1,2
    \\
    -
    \frac{
      \i
      \epsilon^{3\nu\rho\sigma}
      \tilde{n}_\nu
      \tilde{P}_\rho \tilde{\varDelta}_\sigma
    }{(\tilde{P}\cdot\tilde{n})}
    &=
    \frac{\i (\tilde{\bm{P}} \times \tilde{\bm{\varDelta}}_\perp)\cdot\hat{z}}{\tilde{P}_z}
    \\
    \i
    \epsilon^{\alpha\beta\gamma\delta}
    \tilde{n}_\alpha \tilde{\bar{n}}_\beta
    \tilde{\varDelta}_\gamma (\sigma_\delta)_{\lambda'\lambda}
    &=
    - \i
    (\bm{\sigma}_{\lambda'\lambda} \times \tilde{\bm{\varDelta}}_\perp)\cdot\hat{z}
    \,.
  \end{align}
\end{subequations}


\section{Uniqueness of light front synchronization}
\label{sec:unique}

We prove two statements in this appendix.
Firstly,
for any coordinate system utilizing Cartesian spatial coordinates
and a locally time-independent but otherwise arbitrary synchronization convention,
it is impossible to express the physical electromagnetic current density
of a spin-zero target
as a three-dimensional internal density smeared by the probability current.
In other words, there is no function $\rho(\bm{b})$ that is independent of the
state $|\varPsi\rangle$ such that:
\begin{align}
  \langle \varPsi | \hat{j}^\mu(x) | \varPsi \rangle
  =
  \int \d^3 \bm{R} \,
  \Psi^*(\bm{R},t)
  \i \overleftrightarrow{\partial}^\mu
  \Psi(\bm{R},t)
  \,
  \rho(\bm{x}-\bm{R})
\end{align}
for every state $|\varPsi\rangle$.
Secondly,
the only possible coordinate systems which allow a two-dimensional reduction
of the physical current density to be expressed as a two-dimensional internal
density smeared by the probability current is a tilted light front coordinate system
with an arbitrary unit vector $\bm{n}$ defining the longitudinal direction.

One clarification is in order before proceeding.
We mean that a time synchronization convention is
locally time-independent if the proper time element
$\d\tau = \sqrt{g_{\mu\nu}\d x^\mu \d x^\nu}$ measured by a stationary clock
at any location is always $\d t$.
This means we must have $\tilde{g}_{00} = 1$ in the transformed coordinate system.
In other words, we choose \emph{by convention}~\cite{reichenbach2012philosophy}
that the local time coordinate should match the physical time measured
by a local stationary clock.

The coordinate system for the hypothetical time synchronization convention
can be written:
\begin{subequations}
  \label{eqn:B:tilted}
  \begin{align}
    \tilde{x}^0 &= x^0 + S(x) \\
    \tilde{x}^1 &= x^1 \\
    \tilde{x}^2 &= x^2 \\
    \tilde{x}^3 &= x^3
    \,,
  \end{align}
\end{subequations}
where we use tildes for the transformed coordinates,
and where we call $S(x)$ the synchronization function.
The requirement of local time independence imposes:
\begin{align}
  \tilde{g}_{00}
  =
  \frac{\partial x^\mu}{\partial \tilde{x}^0}
  \frac{\partial x^\nu}{\partial \tilde{x}^0}
  g_{\mu\nu}
  =
  \left(1 + \frac{\partial S(x)}{\partial x^0}\right)^2
  =
  1
  \,,
\end{align}
and thus:
\begin{align}
  S(x)
  =
  S(\bm{x})
  \,.
\end{align}
The synchronization function is therefore a function only of space.


\subsection{The non-existence of three-dimensional densities}

To proceed, require an expression for the physical density
$\langle \varPsi | \hat{j}^\mu(\tilde{x}) | \varPsi \rangle$
in terms of the transformed coordinates.
To avoid any complications related to trying to quantize
the theory on a curvilinear equal-time surface,
we proceed within instant form quantization and Minkowski coordinates,
and obtain a formula to which the coordinates can be transformed
according to Eq.~(\ref{eqn:B:tilted}) directly.

Since matrix elements for momentum kets can be written:
\begin{align}
  \langle \bm{p}' | \hat{j}^\mu(0) | \bm{p} \rangle
  =
  2 P^\mu F(\varDelta^2)
  \,,
\end{align}
the physical electromagnetic four-current density is:
\begin{align}
  \langle j^\mu(x) \rangle
  \equiv
  \langle \varPsi | \hat{j}^\mu(x) | \varPsi \rangle
  =
  \int \frac{\d^3\bm{p}}{2E_{\bm{p}}(2\pi)^3}
  \int \frac{\d^3\bm{p}'}{2E_{\bm{p}'}(2\pi)^3}
  \langle \varPsi | \bm{p}' \rangle
  e^{\i p'\cdot x}
  2 P^\mu F(\varDelta^2)
  \langle \bm{p} | \varPsi \rangle
  e^{-\i p\cdot x}
  \,.
\end{align}
Similar to the procedure in Sec.~\ref{sec:derive},
one can substitute $2P^\mu \mapsto \i \overleftrightarrow{\partial}^\mu$ here.
Additionally, the substitution
\begin{align*}
  \varDelta
  \mapsto
  -\i\big( \overrightarrow{\partial} + \overleftarrow{\partial} \big)
\end{align*}
can be made.
If this substitution is formally made within the form factor, we obtain:
\begin{align}
  \langle j^\mu(x) \rangle
  =
  F(-\partial^2)
  \left[
    \left(
    \int \frac{\d^3\bm{p}'}{2E_{\bm{p}'}(2\pi)^3}
    \langle \varPsi | \bm{p}' \rangle
    e^{\i p'\cdot x}
    \right)
    \i \overleftrightarrow{\partial}^\mu
    \left(
    \int \frac{\d^3\bm{p}}{2E_{\bm{p}}(2\pi)^3}
    \langle \bm{p} | \varPsi \rangle
    e^{-\i p\cdot x}
    \right)
    \right]
  \,.
\end{align}
The position-representation wave function can be defined in a covariant way
in terms of the field operator $\hat{\phi}(x)$:
\begin{align}
  \varPsi(\bm{x},t)
  =
  \langle 0 | \hat{\phi}(x) | \varPsi \rangle
  \,,
\end{align}
which is related to the momentum-representation wave function by:
\begin{align}
  \varPsi(\bm{x},t)
  =
  \int \frac{\d^3\bm{p}}{2E_{\bm{p}}(2\pi)^3}
  \langle \bm{p} | \varPsi \rangle
  e^{-\i p\cdot x}
  \,.
\end{align}
Thus, the physical electromagnetic current density can be written:
\begin{align}
  \langle j^\mu(x) \rangle
  =
  F(-\partial^2)
  \left[
    \varPsi^*(\bm{x},t)
    \i \overleftrightarrow{\partial}^\mu
    \varPsi(\bm{x},t)
    \right]
  \equiv
  F(-\partial^2)
  \mathscr{P}^\mu(x)
  \,.
\end{align}
Each piece of this expression can be rewritten in terms of the alternative
coordinate system by employing Eq.~(\ref{eqn:B:tilted}):
\begin{align}
  \label{eqn:B:density}
  \langle \tilde{j}^\mu(\tilde{x}) \rangle
  =
  F(-{\tilde{\partial}^2})
  \left[
    \varPsi^*(\tilde{\bm{x}},\tilde{\tau})
    \i \overleftrightarrow{\tilde{\partial}}^\mu
    \varPsi(\tilde{\bm{x}},\tilde{\tau})
    \right]
  \equiv
  F(-{\tilde{\partial}^2})
  \tilde{\mathscr{P}}^\mu(\tilde{x})
  \,.
\end{align}
To proceed, we need the d'Alembertian in the transformed coordinate system.
In a general coordinate system~\cite{nakahara2003geometry}:
\begin{align}
  \tilde{\partial}^2 f(\tilde{x})
  =
  \frac{1}{\sqrt{-\tilde{g}}}
  \tilde{\partial}_\mu
  \Big[
    \sqrt{-\tilde{g}} \tilde{g}^{\mu\nu} \tilde{\partial}_\nu f(\tilde{x})
    \Big]
  \,,
\end{align}
where $\tilde{g} = \det(\tilde{g}_{\mu\nu})$.
The inverse metric in this coordinate system is:
\begin{align}
  \tilde{g}^{\mu\nu}
  &=
  \frac{\partial \tilde{x}^\mu}{\partial x^\alpha}
  \frac{\partial \tilde{x}^\nu}{\partial x^\beta }
  g^{\alpha\beta}
  =
  \left[
    \begin{array}{cccc}
      1 - (\bm{\nabla}S)^2 & -\nabla_xS & -\nabla_yS & -\nabla_zS \\
      -\nabla_xS & -1 & \phantom{-}0 & \phantom{-}0 \\
      -\nabla_yS & \phantom{-}0 & -1 & \phantom{-}0 \\
      -\nabla_zS & \phantom{-}0 & \phantom{-}0 & -1
    \end{array}
    \right]
  \,,
\end{align}
and its determinant is $-1$, meaning the determinant of $\tilde{g}_{\mu\nu}$
is also $-1$.
Thus, the d'Alembertian in the transformed coordinates can be written:
\begin{align}
  \label{eqn:B:alembert}
  \tilde{\partial}^2
  =
  \big(1-(\bm{\nabla}S)^2\big) \tilde{\partial}_0^2
  - (\nabla^2S) \tilde{\partial}_0
  - (\bm{\nabla}S)\cdot\tilde{\bm{\nabla}}\tilde{\partial}_0
  - \tilde{\bm{\nabla}}^2
  \,.
\end{align}
If a
distribution $\rho(\tilde{\bm{b}})$ defining the
three-dimensional state-independent electric charge density exists,
then the following relation should be satisfied for any state $|\varPsi\rangle$:
\begin{align}
  \label{eqn:false}
  \langle \tilde{j}^\mu(\tilde{x}) \rangle
  =
  \int \d^3\tilde{\bm{R}} \,
  \tilde{\mathscr{P}}^\mu(\tilde{\bm{R}},\tilde{\tau})
  \rho(\tilde{\bm{x}}-\tilde{\bm{R}})
  \,.
\end{align}
We shall proceed to show that no such distribution exists,
by means of a proof by contradiction.

To start, we note that for Eq.~(\ref{eqn:false}) to hold,
the integral on the right-hand side must converge,
which places restrictions on the distribution $\rho(\tilde{\bm{b}})$
which depend on the restrictions in place on the probability four-current
$\tilde{\mathscr{P}}^\mu(\tilde{\bm{R}},\tilde{\tau})$.
A reasonable restriction on the probability current is that
$\tilde{\mathscr{P}}^\mu(\tilde{\bm{R}},\tilde{\tau})$ be a Schwartz function,
meaning that it and all its derivatives vanish at spatial infinity faster than
any power of $|\tilde{\bm{R}}|$.
In this case, $\rho(\tilde{\bm{b}})$ is a tempered distribution,
a broad space of generalized functions which includes the Dirac delta distribution
and its derivatives.
(See Chapter 9 of Folland~\cite{folland2009fourier} for more details.)
Using Eqs.~(\ref{eqn:B:density}) and (\ref{eqn:B:alembert}),
this means:
\begin{align}
  \label{eqn:B:require}
  F\Big(
  -\big(1-(\bm{\nabla}S)^2\big) \tilde{\partial}_0^2
  + (\nabla^2S) \tilde{\partial}_0
  + (\bm{\nabla}S)\cdot\tilde{\bm{\nabla}}\tilde{\partial}_0
  + \tilde{\bm{\nabla}}^2
  \Big)
  \tilde{\mathscr{P}}^\mu(\tilde{x})
  =
  \int \d^3\tilde{\bm{R}}
  \tilde{\mathscr{P}}^\mu(\tilde{\bm{R}},\tilde{\tau})
  \rho(\tilde{\bm{x}}-\tilde{\bm{R}})
  \,.
\end{align}
Because the Fourier transform of a tempered distribution can always be
defined~\cite{folland2009fourier}, we can
(assuming Eq.~(\ref{eqn:false}) is true) introduce
$\varphi(\bm{k})$ as the Fourier
transform of the charge density:
\begin{align}
  \rho(\tilde{\bm{b}})
  =
  \int \frac{\d^3\bm{k}}{(2\pi)^3}
  \varphi(\bm{k})
  e^{-\i\bm{k}\cdot\tilde{\bm{b}}}
  \,,
\end{align}
it follows that:
\begin{align}
  \int \d^3\tilde{\bm{R}}
  \tilde{\mathscr{P}}^\mu(\tilde{\bm{R}},\tilde{\tau})
  \rho(\tilde{\bm{x}}-\tilde{\bm{R}})
  =
  \varphi(\i\tilde{\bm{\nabla}})
  \tilde{\mathscr{P}}^\mu(\tilde{\bm{x}},\tilde{\tau})
  \,,
\end{align}
and Eq.~(\ref{eqn:B:require}) becomes:
\begin{align}
  \label{eqn:B:3D}
  F\Big(
  -\big(1-(\bm{\nabla}S)^2\big) \tilde{\partial}_0^2
  + (\nabla^2S) \tilde{\partial}_0
  + (\bm{\nabla}S)\cdot\tilde{\bm{\nabla}}\tilde{\partial}_0
  + \tilde{\bm{\nabla}}^2
  \Big)
  \tilde{\mathscr{P}}^\mu(\tilde{x})
  =
  \varphi(\i\tilde{\bm{\nabla}})
  \tilde{\mathscr{P}}^\mu(\tilde{\bm{x}},\tilde{\tau})
  \,.
\end{align}
The requirement that this holds for any state imposes:
\begin{align}
  \label{eqn:B:absurd}
  F\Big(
  -\big(1-(\bm{\nabla}S)^2\big) \tilde{\partial}_0^2
  + (\nabla^2S) \tilde{\partial}_0
  + (\bm{\nabla}S)\cdot\tilde{\bm{\nabla}}\tilde{\partial}_0
  + \tilde{\bm{\nabla}}^2
  \Big)
  =
  \varphi(\i\tilde{\bm{\nabla}})
  \,,
\end{align}
in which the right-hand side depends only on spatial derivatives.
This equality can only hold if the left-hand side does not depend
on $\tilde{\partial}_0$, which imposes the following three constraints
that cannot simultaneously hold:
\begin{subequations}
  \label{eqn:B:conditions}
  \begin{align}
    \nabla^2 S
    &=
    0
    \\
    (\bm{\nabla}S)^2
    &=
    1
    \\
    \bm{\nabla} S
    &=
    0
    \,.
  \end{align}
\end{subequations}
In particular, the second and third conditions contradict each other.
We therefore conclude that
Eq.~(\ref{eqn:false}) is false and thus that
no state-independent three-dimensional
internal charge density exists,
regardless of the synchronization convention used.


\subsection{The uniqueness of light front synchronization}

Although no three-dimensional internal density exists,
we already know that light front synchronization allows
an internal two-dimensional density to be found.
We next inquire whether other synchronization conventions
permit alternative two-dimensional densities to be obtained.
(For instance, is there a synchronization convention that is
rotationally symmetric and allows an angular density to be obtained
by integrating out the radial coordinate?)

In the formulation used in this appendix,
light front synchronization works by eliminating derivatives with
respect to the spatial coordinate that has been integrated out.
Thus, the term $(\bm{\nabla}S)\cdot\tilde{\bm{\nabla}}\tilde{\partial}_0$
in the d'Alembertian can be made to vanish without requiring that
$\bm{\nabla}S=0$.
The way that is works with $S(\bm{x}) = z$ as an example is:
\begin{align}
  \int_{-\infty}^\infty \d\tilde{z} \,
  F\Big(
  \tilde{\partial}_z
  \tilde{\partial}_0
  + \tilde{\bm{\nabla}}^2
  \Big)
  \tilde{\mathscr{P}}^\mu(\tilde{x})
  &=
  \sum_{n=0}^\infty
  \frac{F^{(n)}(0)}{n!}
  \int_{-\infty}^\infty \d\tilde{z} \,
  \Big(
  \tilde{\partial}_z
  \tilde{\partial}_0
  + \tilde{\bm{\nabla}}^2
  \Big)^n
  \tilde{\mathscr{P}}^\mu(\tilde{x})
  \notag \\ &=
  \sum_{n=0}^\infty
  \sum_{k=0}^n
  \frac{n!}{k!(n-k)!}
  \frac{F^{(n)}(0)}{n!}
  \int_{-\infty}^\infty \d\tilde{z} \,
  \tilde{\partial}_z^{n-k}
  \tilde{\partial}_0^{n-k}
  (\tilde{\bm{\nabla}}^2)^k
  \tilde{\mathscr{P}}^\mu(\tilde{x})
  \notag \\ &=
  \sum_{n=0}^\infty
  \frac{F^{(n)}(0)}{n!}
  \int_{-\infty}^\infty \d\tilde{z} \,
  (\tilde{\bm{\nabla}}^2)^n
  \tilde{\mathscr{P}}^\mu(\tilde{x})
  =
  \int_{-\infty}^\infty \d\tilde{z} \,
  F(\tilde{\bm{\nabla}}^2)
  \tilde{\mathscr{P}}^\mu(\tilde{x})
  \,.
\end{align}
In the last line, we have dropped all total derivatives with respect to $z$
that appeared in the second line,
since the associated surface terms at infinity vanish.
Since the mixed derivative in the d'Alembertian has been eliminated by integration,
the third condition in Eq.~(\ref{eqn:B:conditions}) is no longer necessary
to eliminate the time derivative dependence from the form factor.
Thus, to obtain two-dimensional densities, only the first two conditions
of Eq.~(\ref{eqn:B:conditions}) are required to hold.

The most general solution to $\nabla^2S=0$ without a singularity
at the origin is:
\begin{align}
  S(\bm{r})
  =
  \sum_{l=0}^\infty
  r^l
  \sum_{m=-l}^l
  C_{lm} Y_{lm}(\theta,\phi)
  \,.
\end{align}
For this general solution:
\begin{multline}
  (\bm{\nabla}S)^2
  =
  \sum_{l,l'}
  r^{l+l'-2}
  \sum_{m=-l}^{l}
  \sum_{m'=-l'}^{l'}
  C_{lm}
  C_{l'm'}^*
  \bigg(
  ll'
  Y_{lm}(\theta,\phi)
  Y^*_{l'm'}(\theta,\phi)
  \\
  +
  \frac{\partial Y_{lm}(\theta,\phi)}{\partial\theta}
  \frac{\partial Y^*_{l'm'}(\theta,\phi)}{\partial\theta}
  +
  \frac{1}{\sin^2\theta}
  \frac{\partial Y_{lm}(\theta,\phi)}{\partial\phi}
  \frac{\partial Y^*_{l'm'}(\theta,\phi)}{\partial\phi}
  \bigg)
  \,.
\end{multline}
To obtain $(\bm{\nabla}S)^2 = 1$,
it is necessary that every power of $r$ greater than $0$
vanish in this expression.
Thus, for any $n > 0$:
\begin{multline}
  \sum_{l=0}^{n+2}
  \sum_{l'=0}^{n+2}
  \delta_{l',n+2-l}
  \sum_{m=-l}^{l}
  \sum_{m'=-l'}^{l'}
  C_{lm}
  C^*_{l'm'}
  \bigg(
  ll'
  Y_{lm}(\theta,\phi)
  Y^*_{l'm'}(\theta,\phi)
  \\
  +
  \frac{\partial Y_{lm}(\theta,\phi)}{\partial\theta}
  \frac{\partial Y^*_{l'm'}(\theta,\phi)}{\partial\theta}
  +
  \frac{1}{\sin^2\theta}
  \frac{\partial Y_{lm}(\theta,\phi)}{\partial\phi}
  \frac{\partial Y^*_{l'm'}(\theta,\phi)}{\partial\phi}
  \bigg)
  =
  0
  \,.
\end{multline}
Integrating this over the surface of a unit sphere,
and using integration by parts, gives:
\begin{multline}
  \sum_{l=0}^{n+2}
  \sum_{l'=0}^{n+2}
  \delta_{l',n+2-l}
  \sum_{m=-l}^{l}
  \sum_{m'=-l'}^{l'}
  C_{lm}
  C^*_{l'm'}
  \int_0^\pi \d\theta \sin\theta
  \int_0^{2\pi} \d\phi \,
  Y^*_{l'm'}(\theta,\phi)
  \bigg(
  ll'
  Y_{lm}(\theta,\phi)
  +
  \\
  -
  \frac{1}{\sin\theta}
  \frac{\partial}{\partial\theta}\left[
    \sin\theta
    \frac{\partial Y_{lm}(\theta,\phi)}{\partial\theta}
    \right]
  -
  \frac{1}{\sin^2\theta}
  \frac{\partial^2 Y_{lm}(\theta,\phi)}{\partial\phi^2}
  \bigg)
  =
  0
  \,.
\end{multline}
Using the differential equation defining the spherical harmonics,
and their orthogonality, gives:
\begin{align}
  \sum_{l=0}^{n+2}
  \sum_{l'=0}^{n+2}
  \delta_{l',n+2-l}
  \sum_{m=-l}^{l}
  \sum_{m'=-l'}^{l'}
  C_{lm}
  C^*_{l'm'}
  \delta_{ll'}
  \delta_{mm'}
  l (l' + l + 1)
  =
  0
  \,.
\end{align}
The Kronecker delta product $\delta_{l',n+2-l} \delta_{ll'}$
will give $0$ automatically if $n$ is odd.
For even $n$,
using the Kronecker deltas to eliminate most the sums gives:
\begin{align}
  \sum_{m=-l}^{l}
  |C_{lm}|^2
  \bigg|_{l = n/2 + 1}
  =
  0
  \,.
\end{align}
Since this must be true for even $n > 0$, it must be true for any $l > 1$.
Thus $C_{lm} = 0$ for all $l>1$.
The most general possible $S$ satisfying the first two conditions
of Eq.~(\ref{eqn:B:conditions}) is thus:
\begin{align}
  S(\bm{r})
  =
  S_0
  +
  \bm{n}\cdot\bm{r}
  \,,
\end{align}
with $S_0$ a constant and $\bm{n}^2 = 1$.
Here $S_0$ is simply a global time offset.
This most general possible solution defines light front synchronization,
with $\bm{n}$ defining the longitudinal direction.
Therefore, light front synchronization is the only synchronization convention
that allows an internal two-dimensional density to be defined.


\bibliography{references.bib}

\end{document}